\documentclass[journal,10pt]{IEEEtran}
\usepackage{graphicx}
\usepackage{subfigure}
\usepackage{float}
\usepackage{subfloat}
\usepackage{amsmath}
\usepackage{amssymb}
\usepackage{amsthm}
\usepackage{epstopdf}
\usepackage{cases}
\usepackage{color}
\usepackage{makecell}
\usepackage{cite}
\usepackage{algorithmic}
\usepackage[ruled]{algorithm2e}

\newtheorem{Theo}{Theorem}

\usepackage[normalem]{ulem}

\ifCLASSINFOpdf
\else
\fi
\usepackage{caption}
\usepackage{mathtools}

\begin{document}

%
\title{Downlink and Uplink Cooperative Joint Communication and Sensing}
%
%
%

\author{Xu Chen,~\IEEEmembership{Student Member,~IEEE,}
	Zhiyong Feng,~\IEEEmembership{Senior Member,~IEEE,}
	J. Andrew Zhang,~\IEEEmembership{Senior Member,~IEEE,}\\
	Zhiqing Wei,~\IEEEmembership{Member,~IEEE,}
	Xin Yuan,~\IEEEmembership{Member,~IEEE,}
	and Ping Zhang,~\IEEEmembership{Fellow,~IEEE}
	\thanks{}
	\thanks{Xu Chen, Z. Feng, and Z. Wei are with Beijing University of Posts and Telecommunications, Key Laboratory of Universal Wireless Communications, Ministry of Education, Beijing 100876, P. R. China (Email:\{chenxu96330, fengzy, weizhiqing\}@bupt.edu.cn).}
	\thanks{J. A. Zhang is with the Global Big Data Technologies Centre, University of Technology Sydney, Sydney, NSW, Australia (Email: Andrew.Zhang@uts.edu.au).}
	\thanks{Ping Zhang is with Beijing University of Posts and Telecommunications, State Key Laboratory of Networking and Switching Technology, Beijing 100876, P. R. China (Email: pzhang@bupt.edu.cn).}
	\thanks{X. Yuan is with Commonwealth Scientific and Industrial Research Organization (CSIRO), Australia (email: Xin.Yuan@data61.csiro.au).}
	\thanks{Corresponding author: Zhiyong Feng}
}

%
%

\markboth{}%
{Shell \MakeLowercase{\textit{et al.}}: Bare Demo of IEEEtran.cls for IEEE Journals}
%



\maketitle

\newcounter{mytempeqncnt}
\setcounter{mytempeqncnt}{\value{equation}}
\begin{abstract}

Downlink (DL) and uplink (UL) joint communication and sensing (JCAS) technologies have been individually studied for realizing sensing using DL and UL communication signals, respectively. Since the spatial environment and JCAS channels in the consecutive DL and UL JCAS time slots are generally unchanged, DL and UL JCAS can be jointly designed to achieve better sensing and communication performance. In this paper, we propose a novel DL and UL cooperative (DUC) JCAS scheme, including a unified multiple signal classification (MUSIC)-based JCAS sensing scheme for both DL and UL JCAS and a DUC JCAS fusion method. The unified MUSIC JCAS sensing scheme can accurately estimate angle-of-arrival (AoA), range, and Doppler based on a unified MUSIC-based sensing module. The DUC JCAS fusion method can distinguish between the sensing results of the communication user and other dumb targets. Moreover, by exploiting the channel reciprocity, it can also improve the sensing and channel state information (CSI) estimation accuracy. Extensive simulation results validate the proposed DUC JCAS scheme. It is shown that the minimum location and velocity estimation mean square errors of the proposed DUC JCAS scheme are about 20 dB lower than those of the state-of-the-art separated DL and UL JCAS schemes.

\end{abstract}


%
\IEEEpeerreviewmaketitle

\section{Introduction}
%
%
%
%
\subsection{Existing Research and Motivations}
Wireless communication and sensing capabilities are both indispensable for the 6th generation (6G) machine-type applications, e.g., intelligent vehicular networks, manufacturing, and smart cities~\cite{Saad2020, Feng2021JCSC}. Unfortunately, the proliferation of wireless sensing and communication nodes has resulted in severe spectrum congestion problems~\cite{liu2020joint}. In this context, the joint communication and sensing (JCAS) technology has emerged as one of the most promising 6G key technologies. It aims to achieve wireless sensing and communication abilities simultaneously using unified spectrum and transceivers, sharing the same transmitted signals~\cite{Chen2020}. 

Up to now, downlink (DL) and uplink (UL) JCAS utilized in mobile networks have been widely studied, adapting to the DL and UL transmission modes of communication systems, respectively. Sturm \textit{et al}.~\cite{Sturm2011Waveform} proposed an orthogonal frequency-division multiplexing (OFDM)-based JCAS signal processing method, which satisfies both the active range detection and communication requirements by using the echoes of communication signals. Zhang \textit{et al.}~\cite{Zhang2019JCRS} proposed a practical OFDM time-division-duplex (TDD) multibeam scheme to achieve JCAS that is suitable for DL echo sensing, which complies with the prevalent terrestrial packet communication system. As pointed out in~\cite{Andrew2021PMN}, the critical enabler for implementing DL JCAS is the full-duplex (FD) operation to transmit JCAS signals and receive reflections simultaneously.
Seyed Ali~\textit{et al.}~\cite{IBFDJCR} realized an FD JCAS platform that detects targets while communicating with another node by canceling the self-leakage interference with analog and digital self-leakage canceler. In~\cite{Yuan2021}, the authors studied spatio-temporal power optimization problems for multi-input multi-output (MIMO) DL JCAS system. In~\cite{Nizhitong2021}, the authors proposed a UL JCAS method for perceptive mobile networks, allowing a static user and base station (BS) to form a bi-static system to sense the environment. Liu \textit{et al.}~\cite{Liu2020} proposed that super-resolution sensing method can be used to achieve DL active range and Doppler estimation. {\color{blue} In~\cite{Chen2023JCAS}, the authors proposed a refined multiple signal classification (MUSIC)-based JCAS sensing scheme to achieve accurate estimation of the angle of arrival (AoA), range and velocity.} All the above researches have laid fundamentals for implementing both the DL mono-static JCAS that exploits the echoes of DL signals, and UL bi-static JCAS that utilizes the UL signals.

In the consecutive UL and DL time slots, the spatial parameters of the environment can be treated as unchanged, which leads to the channel reciprocity~\cite{liu2020joint}. The JCAS operations in consecutive UL and DL time slots can be treated as independent estimates of highly correlated sensing parameters. UL and DL JCAS can potentially be jointly processed to improve the communication and sensing performance of the entire JCAS system. However, few studies have studied the cooperation between UL and DL JCAS up to now.

\subsection{Contributions}
Taking advantage of the above potential in achieving cooperation between the UL and DL JCAS processes, we propose a DL and UL cooperative (DUC) JCAS processing scheme for OFDM-based systems, which can improve the sensing accuracy and communication reliability. This scheme consists of a unified DUC super-resolution sensing method and a DUC JCAS data fusion method. The unified DUC super-resolution sensing method can use the same MUSIC-based JCAS sensing module to accurately estimate AoA, range, and velocity for both UL bi-static and DL mono-static sensing. {\color{blue} Specifically, the unknown AoA and range of the user equipment (UE) can be estimated in the UL JCAS, and this prior information can be exploited in the DL JCAS processing via beamforming (BF) for effective interference mitigation between communication and sensing.} The DUC JCAS data fusion method integrates the UL and DL JCAS sensing results to distinguish between the sensing results of the communication user and other dumb scatterers, and to improve the sensing accuracy. Besides, it fuses the UL and DL channel state information (CSI) to improve communication reliability.

The main contributions of this paper are summarized as follows. 
\begin{itemize}
	\item[1.] {\color{blue} We propose a DUC JCAS processing scheme that uses a common refined two-dimensional (2D) MUSIC algorithm for estimating AoA, range, and velocity for both UL and DL sensing. The estimated UE's AoA and range in UL JCAS can offer crucial prior information for JCAS BF in DL JCAS processing, which leads to efficient interference mitigation between DL sensing and communication.}
	
	
	\item[2.] We propose a DUC JCAS fusion method by leveraging the correlation between UL and DL JCAS channels. It can distinguish the sensing results of the communication user from the other dumb targets, and improve the sensing estimation accuracy by integrating the UL and DL JCAS processing results.
	
	\item[3.] We propose a DUC JCAS CSI fusion method that can integrate the CSI estimates from both UL and DL channel estimation to reduce the error in CSI estimation by exploiting the channel reciprocity of mobile networks. 
\end{itemize}

Extensive simulations are conducted to validate the proposed DUC JCAS scheme. The simulation results show that the location and velocity estimation mean square errors (MSEs) of the proposed DUC JCAS scheme are about 20 dB lower than the state-of-the-art separated DL and UL JCAS schemes. The communication CSI enhancement can improve the bit error rate (BER) performance.
The remaining parts of this paper are organized as follows. 
In section \ref{sec:system-model}, we describe the system model for the DUC JCAS scheme. 
Section \ref{sec:JCAS_signal_processing} proposes the DUC JCAS sensing processing scheme.
Section \ref{sec:DUC_JCAS_Data_Fusion} proposes the DUC JCAS fusion method.
In section \ref{sec:Simulation}, the simulation results are presented. 
Section \ref{sec:conclusion} concludes this paper.

\textbf{Notations}: Bold uppercase letters denote matrices (e.g., $\textbf{M}$); bold lowercase letters denote column vectors (e.g., $\textbf{v}$); scalars are denoted by normal font (e.g., $\gamma$); the entries of vectors or matrices are referred to with square brackets, for instance, the $q$th entry of vector $\textbf{v}$ is $[\textbf{v}]_{q}$, and the entry of the matrix $\textbf{M}$ at the $m$th row and $q$th column is ${[\textbf{M}]_{n,m}}$; ${{\bf{U}}_s} = {\left[ {\bf{U}} \right]_{:,{N_1}:{N_2}}}$ means the matrices sliced from the $N_1$th to the $N_2$th columns of $\bf U$; $\left(\cdot\right)^H$, $\left(\cdot\right)^{*}$ and $\left(\cdot\right)^T$ denote Hermitian transpose, complex conjugate and transpose, respectively; ${\left\| {\mathbf{v}}_{k}  \right\|_l}$ represents the ${ \ell}$-norm of ${\mathbf{v}}_{k}$, and $\ell_2$-norm is considered in this paper; ${ \text{eig}}(\bf M)$ is the eigenvalue decomposition of $\bf M$, and $E\left( \cdot \right)$ represents the expectation of random variables; ${\bf M}_1 \in \mathbb{C}^{M\times N}$ and ${\bf M}_2 \in \mathbb{R}^{M\times N}$ are ${M\times N}$ complex-value and real-value matrices, respectively; and $v \sim \mathcal{CN}(m,\sigma^2)$ means $v$ follows a circular symmetric complex Gaussian (CSCG) distribution with mean $m$ and variance $\sigma^2$.

\begin{figure}[!t]
	\centering
	\includegraphics[width=0.37\textheight]{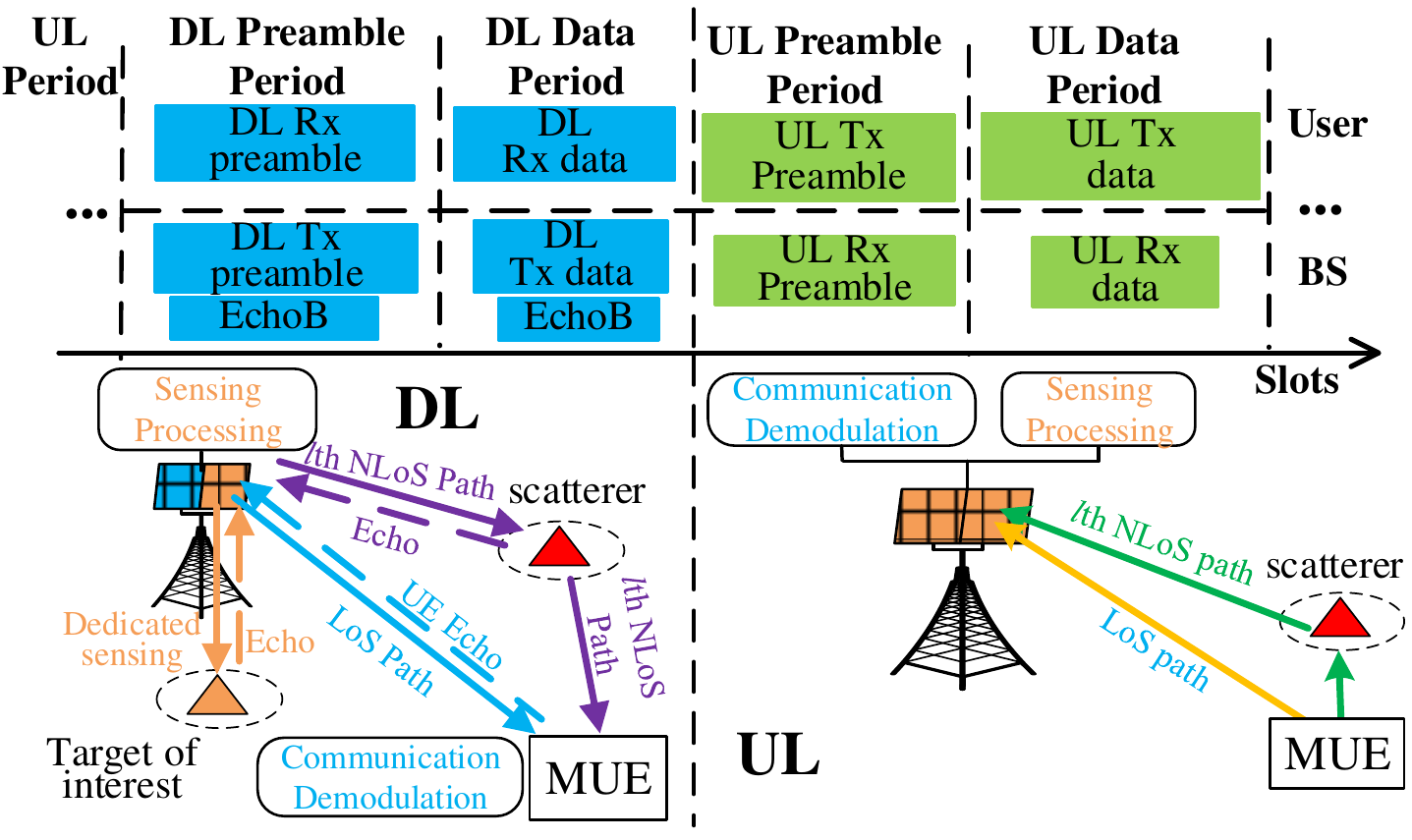}%
	\DeclareGraphicsExtensions.
	\caption{The DUC JCAS scenario.}
	\label{fig: Uplink JCS Model}
\end{figure}
\section{System Model}\label{sec:system-model}
This section presents the DUC JCAS system setup, JCAS channel models, and transmit and received signal models to provide fundamentals for DUC JCAS signal processing.
\subsection{DUC JCAS System Setup} \label{subsec:The Uplink JCS Model}
We consider a DUC JCAS scenario, where the user and BS conduct alternating DL and UL JCAS operations, as shown in Fig.~\ref{fig: Uplink JCS Model}. {\color{blue} A multipath channel model is considered for the block fading channel.} Millimeter-wave (mmWave) and uniform plane arrays (UPAs) are used for the DUC JCAS system. The BS is equipped with two spatially well-separated UPAs and self-leakage canceler to  transmit JCAS
signals and receive reflections simultaneously, as developed in~\cite{IBFDJCR}. {\color{blue} Moreover, synchronization between the BS and user is achieved via a global clock, such as GPS. The clock between them is assumed to be locked, as discussed in \cite{Zhang2022ISAC}. Thus, the timing and carrier frequency residual offset are neglected in the signal model.}

In the UL preamble (ULP) period, the user transmits the ULP signal, and BS receives it for both UL communication setting such as channel estimation and estimating the user's sensing parameters in a bi-static manner. In the UL data (ULD) period, the BS receives and demodulates the ULD signal. In the DL preamble (DLP) period, the user receives the DLP signal from BS for synchronization and channel estimation. BS does not operate JCAS in the DLP period to ensure the best channel estimation. In the DL data (DLD) period, BS transmits the DLD signal to the user and sensing probe signal to the direction-of-interest (DoI), and simultaneously receives the echo signals from both the direction-of-user (DoU) and DoI to perform mono-static sensing. 

As the network environment is generally unchanged for consecutive UL and DL time slots, the UL and DL JCAS can cooperate to enhance both communication and sensing.
DL JCAS is capable of sensing both communication users and the dumb scatterers, while UL JCAS can well estimate the user's sensing parameters as the line-of-sight (LoS) path dominates the mmWave JCAS channel.
As a consequence, after BS performs a round of DL and UL JCAS, it can merge the results to distinguish between the user and the dumb targets, thus improving the sensing performance. Moreover, the UL and DL CSI estimation results can also be merged to improve the communication performance.

Next, we introduce the JCAS transmit signal model and then demonstrate the JCAS channel models.

\subsection{JCAS Transmit Signal} \label{subsec:DUC_signal model}
The UL and DL signals adopt OFDM signals to accommodate the prevalent wireless communication networks. The general OFDM JCAS signal is defined as
\begin{equation}\label{equ:DUC_signal}
	{s^i}( t ) = \sum\limits_{m = 0}^{M_s^i - 1} {\sum\limits_{n = 0}^{N_c^i - 1} {\sqrt {P_t^i} d_{n,m}^i{e^{j2\pi ( {{f_c} + n\Delta {f^i}} )t}}} } {\mathop{\rm Rect}\nolimits} \left(\frac{{t - mT_s^i}}{{T_s^i}}\right),
\end{equation}
where $i = U$ or $D$ are for UL and DL JCAS signals, respectively; ${P_t^i}$ is the transmit power, ${M_s}^i$ and ${N_c^i}$ are the numbers of OFDM symbols and subcarriers for each JCAS process, respectively; ${d_{n,m}^i}$ is the transmit OFDM baseband symbol of the $m$th OFDM symbol of the $n$th subcarrier, $f_c$ is the carrier frequency, $\Delta f^i$ is the subcarrier interval, ${T_s^i} = \frac{1}{{\Delta f^i}} + {T_g^i}$ is the time duration of each OFDM symbol, and ${T_g^i}$ is the guard interval.

\begin{figure}[!t]
	\centering
\includegraphics[width=0.25\textheight]{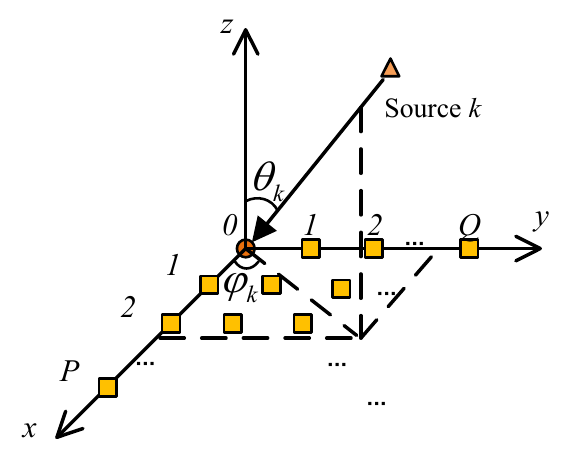}%
	\DeclareGraphicsExtensions.
	\caption{The UPA model.}
	\label{fig: UPA model}
\end{figure}

\subsection{UPA Model} \label{subsec:UPA model}
Fig.~\ref{fig: UPA model} demonstrates the UPA model. The uniform interval between neighboring antenna elements is denoted by $d_a$. The size of UPA is ${P} \times {Q}$. The AoA for receiving or the angle-of-departure (AoD) for transmitting the $k$th far-field signal is ${{\bf{p}}_k} = {( {{\varphi _k},{\theta _k}} )^T}$, where ${\varphi _k}$ is the azimuth angle, and ${\theta _k}$ is the elevation angle. The phase difference between the ($p$,$q$)th antenna element and the reference antenna element is 
\begin{equation}\label{equ:phase_difference}
	{a_{p,q}}\left( {{{\bf{p}}_k}} \right) \! =\! \exp [ { - j\frac{{2\pi }}{\lambda }{d_a}( {p\cos {\varphi _k}\sin {\theta _k} \!+\! q\sin {\varphi _k}\sin {\theta _k}} )} ],
\end{equation}
where $\lambda = c/f_c$ is the wavelength of the carrier, $f_c$ is the carrier frequency, and $c$ is the speed of light. The steering vector for the array is given by 
\begin{equation}\label{equ:steeringVec}
	{\bf{a}}\left( {{{\bf{p}}_k}} \right) = [ {{a_{p,q}}\left( {{{\bf{p}}_k}} \right)} ]\left| {_{p = 0,1,\cdots,P - 1;q = 0,1,\cdots,Q - 1}}\right.,
\end{equation}
where $\mathbf{a}(\mathbf{p}_k) \in \mathbb{C}^{PQ \times 1}$, and ${ {[ {{v_{p,q}}} ]} |_{(p,q) \in {\bf{S}}1 \times {\bf{S}}2}}$ denotes the vector stacked by values ${v_{p,q}}$ satisfying $p\in{\bf{S}}1$ and $q\in{\bf{S}}2$. The steering matrix for $L$ far-field signals is then given by 
\begin{equation}\label{equ:steering Matrix}
	{\bf{A}} = [ {{\bf{a}}( {{{\bf{p}}_1}} ),{\bf{a}}( {{{\bf{p}}_2}} ),\cdots,{\bf{a}}( {{{\bf{p}}_L}} )} ],
\end{equation}
where ${\bf{A}} \in \mathbb{C}^{PQ \times L}$. Then, we demonstrate the UL and DL JCAS channel models. The sizes of the antenna arrays of the BS and the users are ${P_t} \times {Q_t}$ and ${P_r} \times {Q_r}$, respectively.

\subsection{JCAS Channel Models} \label{subsec:JCAS_channel}
BS estimates the Doppler and range from the UL communication channel using bi-static sensing. Therefore, we name it the UL JCAS channel in this paper. Due to the channel reciprocity, the DL communication channel is the transpose of the UL JCAS channel. The DL echo sensing channel consists of the echo path from UE as a scatterer, and the echo paths from other dumb scatterers, as shown in Fig.~\ref{fig: Uplink JCS Model}. Since the signals after multiple reflections are much smaller than those with only one reflection, we only consider echoes directly reflected from scatterers. 

Next, we present the expressions for the aforementioned JCAS channels.

\subsubsection{UL JCAS Channel Model}
The UL JCAS channel response matrix at the $n$th subcarrier of the $m$th OFDM symbol is given by
\begin{equation}\label{equ:H_c^U}
	{\bf{H}}_{C,n,m}^U{\rm{ = }}\sum\limits_{l = 0}^{L - 1} {\left[ \begin{array}{l}
			{b_{C,l}}{e^{j2\pi \left( {{f_{c,d,l}}} \right)mT_s^U}}{e^{ - j2\pi n\Delta {f^U}\left( {{\tau _{c,l}}} \right)}}\\
			\times {\bf{a}}( {{\bf{p}}_{RX,l}^U} ){{\bf{a}}^T}( {{\bf{p}}_{TX,l}^U} )
		\end{array} \right]},
\end{equation}
where ${\bf{H}}_{C,n,m}^U \in \mathbb{C}^{{P_t}{Q_t} \times {P_r}{Q_r}}$, $l = 0$ is for the channel response of the LoS path, and $l \in \{1, \cdots, L-1\}$ is for the paths involved with the $l$th scatterer; ${\bf{a}}( {{\bf{p}}_{RX,l}^U} ) \in \mathbb{C}^{{P_t}{Q_t}} \times 1$ and ${\bf{a}}( {{\bf{p}}_{TX,l}^U} ) \in \mathbb{C}^{{P_r}{Q_r}} \times 1$ are the steering vectors for UL receiving and transmission, respectively; ${\bf{p}}_{RX,l}^U$ and ${\bf{p}}_{TX,l}^U$ are the corresponding AoA and AoD, respectively; ${f_{c,d,0}} = \frac{{{v_{0}}}}{\lambda }$ and ${\tau _{c,0}} = \frac{{{r_{0,1}}}}{c}$ are the Doppler shift and time delay between the user and BS of the LoS path, respectively, with ${v_{0}}$ and ${r_{0,1}}$ being the corresponding radial relative velocity and the distance, respectively; ${f_{c,d,l}} = {f_{d,l,1}} + {f_{d,l,2}}$ and ${\tau _{c,l}} = {\tau _{c,l,1}} + {\tau _{c,l,2}}$ are the aggregate Doppler shift and time delay of the $l$th NLoS path, respectively; ${f_{d,l,1}} = \frac{{{v_{r,l,1}}}}{\lambda }$ and ${f_{d,l,2}} = \frac{{{v_{r,l,2}}}}{\lambda }$ are the Doppler shifts between the user and the $l$th scatterer, and between the $l$th scatterer and the BS, respectively, with ${v_{r,l,1}}$ and ${v_{r,l,2}}$ being the corresponding radial velocities; ${\tau _{c,l,1}} = \frac{{{r_{l,1}}}}{c}$ and ${\tau _{c,l,2}} = \frac{{{r_{l,2}}}}{c}$ are the time delays between the user and the $l$th scatterer, and between BS and the $l$th scatterer, respectively, with ${r_{l,1}}$ and ${r_{l,2}}$ being the corresponding distances. Moreover, ${b_{C,0}} = \sqrt {\frac{{{\lambda ^2}}}{{{{(4\pi {r_{0,1}})}^2}}}}$ and ${b_{C,l}} = \sqrt {\frac{{{\lambda ^2}}}{{{{\left( {4\pi } \right)}^3}{r_{l,1}}^2{r_{l,2}}^2}}} {\beta _{C,l}}$ are the attenuation of the LoS and NLoS paths, respectively; ${\beta _{C,l}}$ is the reflecting factor of the $l$th scatterer, following $\mathcal{CN}(0,\sigma _{C\beta ,l}^2)$~\cite{Rodger2014principles}. 

\begin{figure*}[!t]
	\centering
	\includegraphics[width=0.95\textwidth]{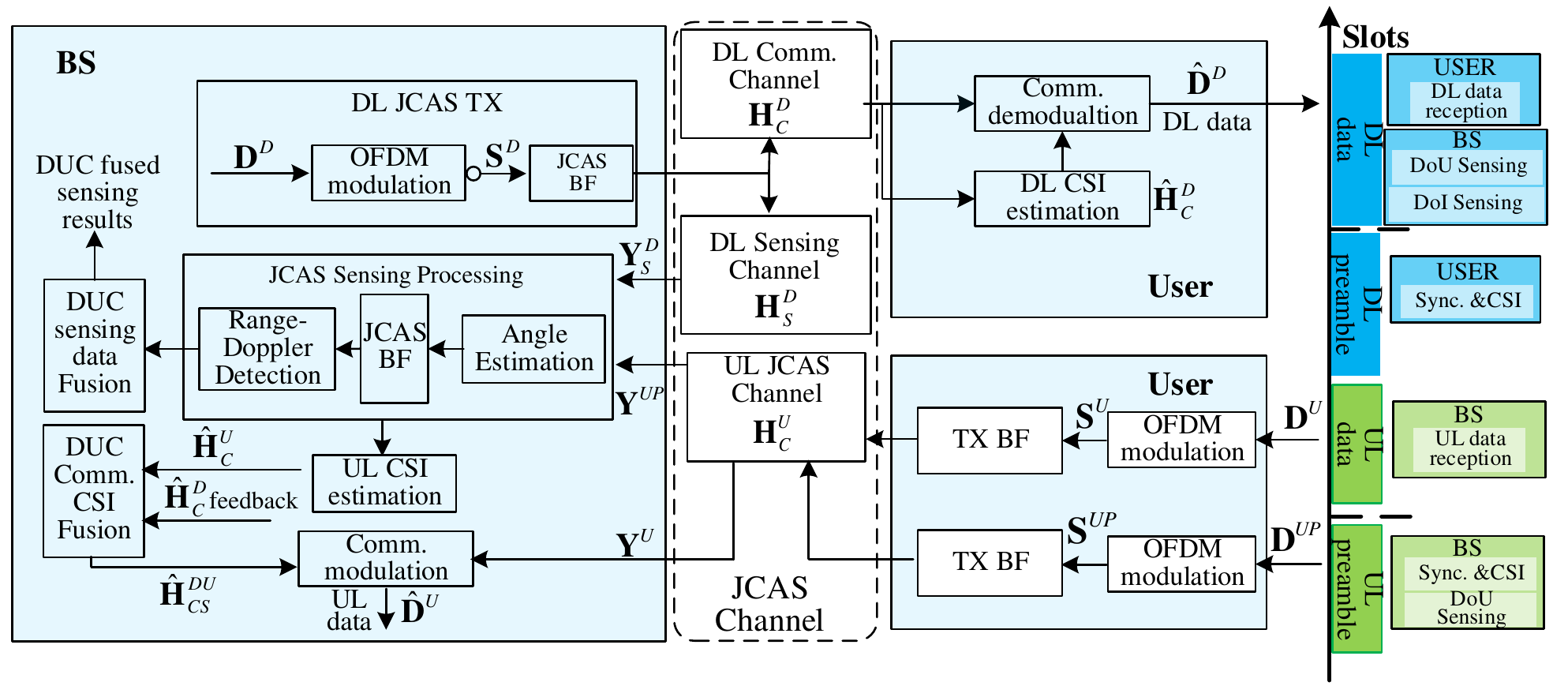}%
	\DeclareGraphicsExtensions.
	\caption{The illustration of DUC JCAS signal processing.}
	\label{fig:DUC_JCS_SignalProcessing}
\end{figure*}

\subsubsection{DL Communication Channel Model}
The DL and UL JCAS adopt the same subcarrier interval and number. Due to the channel reciprocity, the DL communication channel response is the transpose of the UL communication channel response and is given by
\begin{equation}\label{equ:H_C^D}
	{\bf{H}}_{C,n,m}^D{\rm{ = }}\sum\limits_{l = 0}^{L - 1} {\left[ \begin{array}{l}
			{b_{C,l}}{e^{j2\pi \left( {{f_{c,d,l}}} \right)mT_s^D}}{e^{ - j2\pi n\Delta {f^D}\left( {{\tau _{c,l}}} \right)}}\\
			\times {\bf{a}}( {{\bf{p}}_{RX,l}^D} ){{\bf{a}}^T}( {{\bf{p}}_{TX,l}^D} )
		\end{array} \right]},
\end{equation}
where ${\bf{H}}_{C,n,m}^D \in \mathbb{C}^{{P_r}{Q_r} \times {P_t}{Q_t}}$, ${\bf{a}}( {{\bf{p}}_{TX,l}^D} ) \in \mathbb{C}^{{P_t}{Q_t}} \times 1$ and ${\bf{a}}( {{\bf{p}}_{RX,l}^D} ) \in \mathbb{C}^{{P_r}{Q_r}} \times 1$, and ${\bf{p}}_{RX,l}^D = {\bf{p}}_{TX,l}^U$ and ${\bf{p}}_{TX,l}^D = {\bf{p}}_{RX,l}^U$ are the DL communication AoA and AoD, respectively.

\subsubsection{DL Echo Sensing Channel Model}
The response of the DL echo sensing channel at the $n$th subcarrier of the $m$th OFDM symbol is given by
\begin{equation}\label{equ:JCS sensing channel}
	{\bf{H}}_{S,n,m}^D{\rm{ = }}\sum\limits_{l = 0}^{L - 1} {\left[ \begin{array}{l}
			{b_{S,l}}{e^{j2\pi {f_{s,l,1}}mT_s^D}}{e^{ - j2\pi n\Delta {f^D}\left( {{\tau _{s,l}}} \right)}}\\
			\times {\bf{a}}( {{\bf{p}}_{RX,l}^{DS}} ){{\bf{a}}^T}( {{\bf{p}}_{TX,l}^D} )
		\end{array} \right]},
\end{equation}
where ${{\bf{p}}_{TX,l}^D}$ and ${{\bf{p}}_{RX,l}^{DS}}$ are the AoD and AoA of the JCAS transmitter and sensing receive array, respectively; ${{\bf{a}}( {{\bf{p}}_{TX,l}^D})} \in \mathbb{C}^{{P_t}{Q_t} \times 1}$ and ${{\bf{a}}( {{\bf{p}}_{RX,l}^{DS}} )} \in \mathbb{C}^{{P_t}{Q_t} \times 1}$ are the corresponding steering vectors as given in \eqref{equ:steeringVec}. Since the mmWave array is typically small, ${\bf{p}}_{RX,l}^{DS} = {\bf{p}}_{TX,l}^D$. Moreover, ${f_{s,0,1}} = \frac{{2{v_{0}}}}{\lambda }$ and ${f_{s,l,1}} = \frac{{2{v_{r,l,2}}}}{\lambda }$ are the Doppler frequency shifts of the $l$th echo path, with $v_{0}$ and $v_{r,l,2}$ being the corresponding radial relative velocities; ${\tau _{s,0}} = \frac{{2{r_{0,1}}}}{c}$ and ${\tau _{s,l}} = \frac{{2{r_{l,2}}}}{c}$ are the time delays of the $l$th echo path, with $r_{0,1}$ and $r_{l,2}$ being the corresponding ranges; ${b_{S,l}} = \sqrt {\frac{{{\lambda ^2}}}{{{{\left( {4\pi } \right)}^3}{d_{l,2}}^4}}} {\beta _{S,l}}$ with ${\beta _{S,l}}$ being the reflecting factor of the $l$th scatterer that follows $\mathcal{CN}(0,\sigma _{S\beta ,l}^2)$, according to Swerling model~\cite{richards2010principles}. 

\subsection{UL and DL Received JCAS Signals}
\subsubsection{Received Communication Signals}
The received communication signal at the $m$th OFDM symbol of the $n$th subcarrier is 
\begin{equation}\label{equ:y_C^U}
\begin{array}{l}
	{\bf{y}}_{C,n,m}^i = \sqrt {P_t^i} d_{n,m}^i{\bf{H}}_{C,n,m}^i {{\bf{w}}_{TX}^i} + {\bf{n}}_{t,n,m}^i\\
	= \sqrt {P_t^i} d_{n,m}^i\sum\limits_{l = 0}^{L - 1} {\left[ \begin{array}{l}
			{b_{C,l}} \chi _{TX,l}^i {e^{j2\pi mT_s^i{f_{c,d,l}}}}\\
			\times {e^{ - j2\pi n\Delta {f^i}{\tau _{c,l}}}} {\bf{a}}( {{\bf{p}}_{RX,l}^i} )
		\end{array} \right]}  \!\!+\! {\bf{n}}_{t,n,m}^i,
\end{array}
\end{equation}
where $i = U$ and $D$ are for UL and DL JCAS signals, respectively; ${\bf{n}}_{t,n,m}^i$ is the combined noise that contains Gaussian noise and possible reflected interferences, and each element of ${\bf{n}}_{t,n,m}^i$ follows $\mathcal{CN}(0,\sigma _N^2)$; ${\bf{y}}_{C,n,m}^U, {\bf{n}}_{t,n,m}^U \in \mathbb{C}^{{P_t}{Q_t} \times 1}$, ${\bf{y}}_{C,n,m}^D, {\bf{n}}_{t,n,m}^D \in {\mathbb{C}^{{P_r}{Q_r} \times 1}}$; $d_{n,m}^i$ is the transmit symbol, and $P_t^i$ is the transmit power; ${\bf{w}}_{TX}^i$ is the transmit BF vector, and $\chi _{TX,l}^i = {{\bf{a}}^T}( {{\bf{p}}_{TX,l}^i} ) {{\bf{w}}_{TX}^i}$ is the transmit BF gain. We adopt the low-complexity least-square (LS) method to generate ${\bf{w}}_{TX}^i$, i.e., ${\bf{w}}_{TX}^i = {c_0}{[ {{{\bf{a}}^T}( {\tilde{\bf{p}}_{TX,l}^i} )} ]^\dag }$~\cite{Zhang2019JCRS}, where $[\textbf{A}]^\dag$ is the pseudo-inverse matrix of $\textbf{A}$, and ${c_0} = {e^{j2\pi f\phi }}$ is a complex value with unit modulus and arbitrary phase. When the transmit beam is well aligned to the user, ${\tilde{\bf{p}}_{TX,l}^i} \approx {{\bf{p}}_{TX,l}^i}$.

In the ULP and DLP periods, $d_{n,m}^i = \bar d_{n,m}^i$ and $P_t^i = \bar P_t^i$ are the preamble symbols and the corresponding transmit power that are deterministic and known to BS and the user. Without loss of generality, we assume $\bar P_t^i$ is the maximum value of $P_t^i$. The corresponding received signals for preambles are denoted by ${\bar{\bf{y}}}_{C,n,m}^i$. In the DLD and ULD periods, $d_{n,m}^i \in {\Theta _{QAM}}$ is the random data symbol, where ${\Theta _{QAM}}$ is the used quadrature amplitude modulation (QAM) constellation for communication.

\subsubsection{Received Echo Sensing Signal} \label{sec:ULD_SIG}
In the DLD period, the BS can transmit dedicated sensing probe signals to sense the targets in DoI, denoted by ${\bf{p}}_S^D$. Denote ${\bf{w}}_{TX}^{DS}$ to be the transmit BF vector to illuminate direction ${\bf{p}}_S^D$. The echo sensing signal received by BS is expressed as
\begin{equation}\label{equ:y_S^DS}
{\bf{y}}_{S,n,m}^{DS} = {\bf{H}}_{S,n,m}^D\left( \begin{array}{l}
	\sqrt {P_t^D} d_{n,m}^D{\bf{w}}_{TX}^D\\
	+ \sqrt {P_t^{DS}} d_{n,m}^{DS}{\bf{w}}_{TX}^{DS}
\end{array} \right) + {\bf{n}}_{S,n,m}^{DS},
\end{equation}
where ${\bf{y}}_{S,n,m}^{DS} \in \mathbb{C}^{{P_t}{Q_t} \times 1}$, $d_{n,m}^{DS}$ is the dedicated sensing symbol with unit constant modulus, $d_{n,m}^D$ is the DL communication data symbol; $P_t^{DS}$
and $P_t^D$ are the powers for $d_{n,m}^{DS}$ and $d_{n,m}^D$, respectively; and $P_t^{DS} + P_t^D = \bar P_t^D$. Moreover, ${\bf{n}}_{S,n,m}^{DS} \in \mathbb{C}^{{P_t}{Q_t} \times 1}$ is the noisy vector that contains Gaussian noise and possible reflected interferences, with each element following $\mathcal{CN}(0,\sigma _N^2)$. 

From \eqref{equ:JCS sensing channel}, \eqref{equ:y_C^U}, and~\eqref{equ:y_S^DS}, we can see that the UL and DL JCAS paths directly formed by the user and BS contain the identical user's range, radial velocity, and AoA. Therefore, the consecutive UL and DL JCAS can conduct independent estimates of several identical sensing parameters, which is the basis for DUC JCAS processing scheme.

\section{DUC JCAS Sensing Processing}\label{sec:JCAS_signal_processing}
The DUC JCAS processing scheme is shown in Fig.~\ref{fig:DUC_JCS_SignalProcessing}. Specifically, this section demonstrates the unified UL and DL JCAS sensing processing methods, and the JCAS data fusion method will be presented in Section~\ref{sec:DUC_JCAS_Data_Fusion}. {\color{blue} In this section, we first use a unified MUSIC-based module to estimate the AoA, range and Doppler for the UE in the UL JCAS processing, which are then used for the BF in DL JCAS to suppress the mutual interference between DL sensing and communication. Finally, we show that the DL JCAS processing can use the same MUSIC-based module to estimate the ranges and Doppler frequencies of targets as in the UL JCAS processing.}

\subsection{UL JCAS Processing}\label{sec:UplinkJCAS}
In the ULP period, based on \eqref{equ:y_C^U}, the UL CSI estimation at the $n$th subcarrier of the $m$th OFDM symbol is obtained with the LS method as \cite{2010MIMO}
\begin{equation}\label{equ:h_c_U}
	{{{\bf{\hat h}}_{C,n,m}^U = {\bf{\bar y}}_{C,n,m}^U} \mathord{/
			{\vphantom {{{\bf{\hat h}}_{C,n,m}^U = {\bf{\bar y}}_{C,n,m}^U} {\left( {\sqrt {\bar P_t^U} \bar d_{n,m}^U} \right)}}} 
			\kern-\nulldelimiterspace} {( {\sqrt {\bar P_t^U} \bar d_{n,m}^U} )}} \in \mathbb{C}^{P_t Q_t \times 1},
\end{equation}
Since $N_c^U$ subcarriers at $M_s^U$ OFDM preamble symbols are used, we can stack all the CSIs to obtain ${\bf{\hat H}}_C^U \in {\mathbb{C}^{{P_t}{Q_t} \times N_c^UM_s^U}}$, where the $[(m-1)N_c^U + n]$th column of ${\bf{\hat H}}_C^U$ is ${{\bf{\hat h}}_{C,n,m}^U}$. Denote the UL incident signals of $L$ paths as ${\bf{s}}_{n,m}^U \in \mathbb{C}^{L \times 1}$, where
\begin{equation}\label{equ:s_n_m_l}
	\begin{aligned}
{\left[ {{\bf{s}}_{n,m}^U} \right]_l} = \; {b_{C,l}}\chi _{TX,l}^U{e^{j2\pi mT_s^U f_{c,d,l}}} {e^{ - j2\pi n\Delta {f^U}{\tau _{c,l}} }}.
\end{aligned}
\end{equation}
Then, ${\bf{\hat H}}_C^U$ can be expressed as 
\begin{equation}\label{equ:H_C^U}
	{\bf{\hat H}}_C^U = {{\bf{A}}_{U,RX}}{\bf{S}}_{}^U + {\bf{N}}_t^U, 
\end{equation}
where  ${{\bf{A}}_{U,RX}} = { {[ {{\bf{a}}( {{\bf{p}}_{RX,l}^U} )} ]} |_{l = 0,1,...,L - 1}} \in \mathbb{C}^{{P_t}{Q_t} \times L}$ is the steering matrix, and the $[(m-1)N_c^U + n]$th column of ${{\bf{N}}_t^U}$ is ${{{\bf{n}}_{t,n,m}^U} \mathord{/
		{\vphantom {{{\bf{n}}_{t,n,m}^U} {(\sqrt {\bar P_t^U} \bar d_{n,m}^U)}}} 
		\kern-\nulldelimiterspace} {(\sqrt {\bar P_t^U} \bar d_{n,m}^U)}}$. Moreover, ${{\bf{S}}^U} \in \mathbb{C}^{{L \times N_c^UM_s^U}}$ is expressed as
\begin{equation}\label{equ:S_U}
	{{\bf{S}}^U} = {\left. {\left[ {{\bf{s}}_{n,m}^U} \right]} \right|_{(n,m) \in [0,1, \cdots ,N_c^U - 1] \times [0,1, \cdots ,M_s^U - 1]}}.
\end{equation}

\subsubsection{UL JCAS Angle Estimation}
We obtain the AoAs from ${\bf{\hat H}}_C^U$ using the refined MUSIC-based estimation method, as referenced to \cite{Chen2023JCAS}. First, we compute the autocorrelation matrix of ${\bf{Y}}_C^U$ as
\begin{equation}\label{equ:autocorrelation}
	{\bf{R}}_{\bf{x}}^U{\rm{ = }}{{[ {{\bf{\hat H}}_C^U{{( {{\bf{\hat H}}_C^U} )}^H}} ]} \mathord{/
			{\vphantom {{\left[ {{\bf{\hat H}}_C^U{{\left( {{\bf{\hat H}}_C^U} \right)}^H}} \right]} {(M_s^UN_c^U)}}} 
			\kern-\nulldelimiterspace} {(M_s^UN_c^U)}} \in {\mathbb{C}^{{P_t}{Q_t} \times {{P_t}{Q_t}}}}.
\end{equation}
Applying eigenvalue decomposition to ${\bf{R}}_{\bf{x}}^U$, we obtain
\begin{equation}\label{equ:eigenvalue deomposition of Rx}
	[ {{\bf{U}}_x^U,{\bf{\Sigma }}_x^U} ] = \text{eig}\left( {{\bf{R}}_{\bf{x}}^U} \right),
\end{equation}
where $\text{eig}(\bf M)$ represents the eigenvalue decomposition of $\bf M$, ${\bf{\Sigma }}_x^U$ is the real-value eigenvalue diagonal matrix in descending order, and ${\bf{U}}_x^U$ is the orthogonal eigen matrix. The number of incident signals is denoted by $N_x^U$. The noise subspace of ${{\bf{R}}_{\bf{x}}^U}$ is ${\bf{U}}_N^U = {\left[ {{\bf{U}}_x^U} \right]_{:,N_x^U + 1:{P_t}{Q_t}}}$, and then we formulate the angle spectrum function as~\cite{HAARDT2014651}
\begin{equation}\label{equ:angle_spectrum_function}
	f_a^U( {\bf{p}} ) = {{\bf{a}}^H}( {\bf{p}} ){\bf{U}}_N^U{( {{\bf{U}}_N^U} )^H}{\bf{a}}( {\bf{p}} ),
\end{equation}
where ${\bf{a}}\left( {\bf{p}} \right)$ is given in~\eqref{equ:steeringVec}. The angle spectrum is further obtained as~\cite{HAARDT2014651}
\begin{equation}\label{equ:angle_spectrum}
	S_a^U\left( {\bf{p}} \right) = {[ {{{\bf{a}}^H}\left( {\bf{p}} \right){\bf{U}}_N^U{{\left( {{\bf{U}}_N^U} \right)}^H}{\bf{a}}\left( {\bf{p}} \right)} ]^{ - 1}}.
\end{equation}
The minimum points of $f_a^U( {\bf{p}} )$, i.e, the maximum points of $S_a^U\left( {\bf{p}} \right)$ are the estimated AoAs. We then use a 2D two-step Newton descent method in \cite{Chen2023JCAS} to derive the minimum points of $f_a^U( {\bf{p}} )$, which is demonstrated in \textbf{Algorithm~\ref{alg:Two-step_descent}}. The initial points for Newton descent iteration are given by a coarse-granularity grid search \cite{Nizhitong2021}. 

To identify the minimum of $f_a^U( {\bf{p}} )$, we substitute $f( {\bf{p}} )$, ${\bf{H}}_{\bf{p}}\left( {\bf{p}} \right)$, and ${\nabla _{\bf{p}}}f\left( {\bf{p}} \right)$ in \textbf{Algorithm~\ref{alg:Two-step_descent}} with \eqref{equ:angle_spectrum_function}, Hessian matrix and the gradient vector of $f_a^U( {\bf{p}} )$, respectively. Note that \textbf{Algorithm~\ref{alg:Two-step_descent}} can also be used in the one-dimensional (1D) parameter estimation by treating the second parameter to be a constant value.

\begin{algorithm}[!t]
	\caption{2D two-step Newton descent minimum searching method~\cite{Chen2023JCAS}}
	\label{alg:Two-step_descent}
	\KwIn{The range of $\varphi $: ${\Phi _\varphi }$; the range of $\theta $: ${\Phi _\theta }$; the number of initial grid points: ${N_i}$; the maximum iteration round $ind_{max}$; the MUSIC spectrum function: $f(\bf p)$.
	}
	\KwOut{Estimation results: ${\bf{\Theta }}\!\! =\!\! { {\{ {{{{\bf{\hat p}}}_k}} \}} |_{k \in \{ {0,...,{N_s} - 1} \}}}$.}
	\textbf{Initialize: } \\
	1) {\color{blue}${\Phi _\varphi }$ and ${\Phi _\theta }$ are both divided evenly into ${N_i} - 1$ pieces by ${N_i}$ grid points to generate a meshgrid ${\hat \Phi _\varphi } \times {\hat \Phi _\theta}$.}\\
	2) Set a null space ${\bf{\Theta }}$.\\
	\textbf{Process: } \\
	\textbf{Step 1}: \ForEach{${{\bf{p}}_{i,j}} \in {\hat \Phi _\varphi } \times {\hat \Phi _\theta}$}
	{
		Calculate the spatial spectrum as ${\bf{S}}$, where ${[{\bf{S}}]_{i,j}} = [f\left( {{{\bf{p}}_{i,j}}} \right)]^{-1}$.
	}
	\textbf{Step 2}: Search the maximal values of ${\bf{S}}$ to form the set ${\bar \Theta _d}$.
	
	\textbf{Step 3}: Derive the Hessian matrix and the gradient vector of $f( {\bf{p}} )$ as ${\bf{H}}_{\bf{p}}\left( {\bf{p}} \right)$ and ${\nabla _{\bf{p}}}f\left( {\bf{p}} \right)$, respectively. 
	
	\textbf{Step 4}: \ForEach{${{\bf{p}}_{i,j}} \in {\bar \Theta _d}$}{
		\textit{k}=1\;
			${{\bf{p}}^{( 0 )}} = {{\bf{p}}_{i,j}}$\;
			${{\bf{p}}^{( k )}} = {{\bf{p}}^{\left( {k - 1} \right)}} - {[ {{{\bf{H}}_{\bf{p}}}( {{{\bf{p}}^{( {k - 1} )}}} )} ]^{ - 1}}{\nabla _{\bf{p}}}f( {{{\bf{p}}^{( {k - 1} )}}} )$\;
			\While{$\| {{{\bf{p}}^{\left( k \right)}} - {{\bf{p}}^{( {k - 1} )}}} \| > \varepsilon $ and $k \le ind_{max}$}
			{
				${{\bf{p}}^{( k )}}= {{\bf{p}}^{( {k - 1} )}} - {[ {{{\bf{H}}_{\bf{p}}}( {{{\bf{p}}^{( {k - 1} )}}} )} ]^{ - 1}}{\nabla _{\bf{p}}}f( {{{\bf{p}}^{( {k - 1} )}}} )$\;
				
			}
			${{\bf{p}}^{\left( k \right)}}$ is put into output set ${\bf{\Theta }}$\;
	}
\end{algorithm}

By applying \textbf{Algorithm~\ref{alg:Two-step_descent}}, we estimate the AoAs denoted by ${\bf{\Theta}}^U = { {\{ {{{{\bf{\hat p}}}^U_k}} \}} |_{k \in \{ {0,1,...,{N_x^U} - 1} \}}}$. Moreover, ${\bf{\Theta}}^U$ is sorted in the descending order by the value of $S_a^U\left( {{{{\bf{\hat p}}}^U_k}}\right)$. {\color{blue} Typically, $\hat{\bf{p}}_0^U$ is the estimated user's AoA because the LoS path dominates the UL JCAS channel.}

The BF matrix to receive the incident signals can be obtained by solving the following optimization problem:
\begin{equation}\label{equ:W_RxU}
	{\bf{W}}_{RX}^U = \mathop {\arg \min }\limits_{\bf{W}} \| {{{\bf{W}}^H}{{\bf{A}}_{U,RX}}{\bf{S}}^U - {\bf{S}}_{}^U} \|_2^2,
\end{equation}
where the $m$th column of ${\bf{W}}_{RX}^U$ is the BF vector that receives the signal from the $m$th direction in ${\bf{\Theta}}^U$.
Since the problem is convex, we have
\begin{equation}\label{equ:w_RxUResult}
	{\bf{W}}_{RX}^U = {[ {{{\bf{A}}_{U,RX}}{{( {{{\bf{A}}_{U,RX}}} )}^H}} ]^{ - 1}}{{\bf{A}}_{U,RX}}.
\end{equation}
For the estimated steering matrix, i.e., ${{\bf{\tilde A}}_{U,RX}} = { {[ {{\bf{a}}( {{\bf{\hat p}}_k^U} )} ]} |_{k = 0,1,. \cdots ,N_x^U - 1}}$, the received BF matrix is expressed as
\begin{equation}\label{equ:w_RxUResult_est}
	{\bf{\tilde W}}_{RX}^U = {[ {{{{\bf{\tilde A}}}_{U,RX}}{{( {{{{\bf{\tilde A}}}_{U,RX}}} )}^H}} ]^{ - 1}}{{\bf{\tilde A}}_{U,RX}}. 
\end{equation}
Then, we normalize each column of ${\bf{\tilde W}}_{RX}^U$ to obtain ${\bf{\bar W}}_{RX}^U$. Since the LoS path generally has dominant power in mmWave systems, we use the first column of ${\bf{\bar W}}_{RX}^U$, denoted by ${\bf{ w}}_{RX}^U = {[ {{\bf{\bar W}}_{RX}^U} ]_{:,1}}$, to constructively combine the LoS path's signals.

Then, the UL CSI obtained by BS is
\begin{equation}\label{equ:y_CS}
	{\bf{\hat h}}_{CS}^U = {( {{\bf{w}}_{RX}^U} )^H}{\bf{\hat H}}_C^U, 
\end{equation}
where ${\bf{\hat h}}_{CS}^U \in {\mathbb{C}^{1 \times N_c^UM_s^U}}$ also includes the LoS path's sensing parameters. 


\subsubsection{UL JCAS Range-Doppler Estimation} \label{sec:UL_JCAS_range_Doppler}
Reshape ${\bf{\hat h}}_{CS}^U$ to a $N_c^U \times M_s^U$ matrix, denoted by ${\bf{\hat H}}_{CS}^U$, then combine \eqref{equ:s_n_m_l}, \eqref{equ:H_C^U}, \eqref{equ:S_U} and \eqref{equ:y_CS}. We can obtain the $(n,m)$th element of ${\bf{\hat H}}_{CS}^U$ as
\begin{equation}\label{equ:y_CS_nm}
\begin{aligned}
	&\hat h_{CS,n,m}^U = h_{CS,n,m}^U + w_{t,n,m}^U\\
	&= \sum\limits_{l = 0}^{L - 1} {\left[ \begin{array}{l}
			{b_{C,l}}{e^{j2\pi mT_s^U f_{c,d,l}}}{e^{ - j2\pi n\Delta {f^U} \tau _{c,l}}}\\
			\times \chi _{TX,l}^U\varpi _{RX,l}^U
		\end{array} \right]}  + w_{t,n,m}^U
\end{aligned},
\end{equation}
where $w_{t,n,m}^U{\rm{ = }}{\left( {{\bf{w}}_{RX}^U} \right)^H}{{{\bf{n}}_{t,n,m}^U} \mathord{/
{\vphantom {{{\bf{n}}_{t,n,m}^U} {(\sqrt {\bar P_t^U} \bar d_{n,m}^U)}}} 
\kern-\nulldelimiterspace} {(\sqrt {\bar P_t^U} \bar d_{n,m}^U)}}$ is the transformed noise, $h_{CS,n,m}^U = {( {{\bf{w}}_{RX}^U} )^H}{\bf{H}}_{C,n,m}^U {{\bf{w}}_{TX}^U} $ is the actual communication CSI, and $\varpi _{RX,l}^U = {( {\bf{w}}_{RX}^U )^H}{\bf{a}}( {{\bf{p}}_{RX,l}^U} )$ is the gain at the $l$th AoA. Since ${{\bf{w}}_{RX}^U}$ points at the LoS path, $\varpi _{RX,0}^U$ is much larger than $\varpi _{RX,l}^U$ ($l \ne 0$). 

Notice that $\hat h_{CS,n,m}^U$ contains independent complex exponential functions for range and Doppler, i.e., ${e^{j2\pi mT_s^U f_{c,d,l}}}$ and ${e^{ - j2\pi n\Delta {f^U}{{\tau }_{c,l}}}}$. Here, we define the range and Doppler steering vectors, respectively, as
\begin{equation}\label{equ:aUr}
{\bf{a}}_r( {r;\Delta f,{N_c}} ) = { {[ {{e^{ - j2\pi n\Delta f\frac{r}{c}}}} ]} |_{n = 0,1,...,{N_c} - 1}},
\end{equation}
\begin{equation}\label{equ:aUf}
{{\bf{a}}_f}\left( {f;{T_s},{M_s}} \right) = { {[ {{e^{j2\pi m{T_s}f}}} ]} |_{m = 0,1,...,{M_s} - 1}}.
\end{equation}
Then, the UL range and Doppler steering matrices are defined as
\begin{equation}\label{equ:AUr}
	{\bf{A}}_{\bf{r}}^U = { {[ {{{\bf{a}}_r}( {{{ r}_l};\Delta {f^U},N_c^U} )} ]} |_{l = 0,1,...,L - 1}} \in \mathbb{C}^{N_c^U \times L},
\end{equation}
\begin{equation}\label{equ:AUf}
{\bf{A}}_{\bf{f}}^U = { {[ {{{\bf{a}}_f}( { f_{c,d,l};T_s^U,M_s^U} )} ]} |_{l = 0,1,...,L - 1}} \in \mathbb{C}^{M_s^U \times L},
\end{equation}
where ${ r_l} = { \tau_{c,l}}\times c$.

According to \eqref{equ:y_CS_nm}, ${\bf{\hat H}}_{CS}^U$ can be written in matrix form as
\begin{equation}\label{equ:Y_USbar}
	{\bf{\hat H}}_{CS}^U = {\bf{A}}_{\bf{r}}^U{\bf{S}}_{S}^U{\left( {{\bf{A}}_{\bf{f}}^U} \right)^T} + {\bf{W}}_{tr}^U,
\end{equation}
where ${\bf{S}}_{S}^U = diag( {{{ {[ {\sqrt {P_t^U} {b_{C,l}}\varpi _{RX,l}^U\chi _{TX,l}^U} ]} |}_{l = 0, \cdots ,L - 1}}} )$ is irrelevant to ${\bf{A}}_{\bf{r}}^U$ and ${{\bf{A}}_{\bf{f}}^U}$, and ${[ {{\bf{W}}_{tr}^U} ]_{n,m}} = w_{t,n,m}^U$. We use the \textbf{Theorem~\ref{Theo:range_Doppler}} in \cite{Chen2023JCAS}, to estimate the range and Doppler.
\begin{Theo} \label{Theo:range_Doppler}
	{\rm The range and Doppler steering matrices are ${{\bf{A}}_{\bf{r}}} = { {[ {{{\bf{a}}_r}( {{r_l};\Delta f,{N_c}} )} ]} |_{l = 0,1,...,L - 1}}$ and ${{\bf{A}}_{\bf{f}}} = {\left. {\left[ {{{\bf{a}}_f}\left( {{f_l};T,{M_s}} \right)} \right]} \right|_{l = 0,1,...,L - 1}}$. If ${\bf{\hat H}} = {{\bf{A}}_{\bf{r}}}{{\bf{S}}_s}{( {{{\bf{A}}_{\bf{f}}}} )^T} + {\bf{W}} \in \mathbb{C}^{{N_c} \times {M_s}}$, where ${{\bf{S}}_s} = diag( {{{ {[ {{a_l}} ]} |}_{l = 0, \cdots ,L - 1}}} )$, ${a_l}$ is a complex value irrelevant to ${{\bf{A}}_{\bf{r}}}$ and ${{\bf{A}}_{\bf{f}}}$, and ${\bf{W}}$ is zero-mean Gaussian noise matrix. Let the noise subspaces of ${{\bf{\hat H}}}$ and ${ {{\bf{\hat H}}}^T}$ be ${{\bf{U}}_{x,rN}}$ and ${{\bf{U}}_{x,fN}}$, respectively. Then, the minimal values of $\| {{{\bf{U}}_{x,rN}}^H{{\bf{a}}_r}\left( r \right)} \|_2^2$ and $\| {{{\bf{U}}_{x,fN}}^H{{\bf{a}}_f}( f )} \|_2^2$ are $r = {r_l}$ and $f = {f_l}$, respectively~\cite{Chen2023JCAS}.
		\begin{proof}			
			The proof is provided in the \textbf{Appendix C} in \cite{Chen2023JCAS}. 
		\end{proof}
	}
\end{Theo}

{\color{blue} Based on \textbf{Theorem~\ref{Theo:range_Doppler}}, following the same approach as taken in \eqref{equ:autocorrelation}-\eqref{equ:angle_spectrum}, we can obtain the range and Doppler spectra. From the spectra, we use the steps 3-4 in \textbf{Algorithm~\ref{alg:Two-step_descent}} to find the peaks that are the estimates of ranges and Dopper frequencies.}
We first derive the correlation matrices of ${\bf{\hat H}}_{CS}^U$ and ${( {{\bf{\hat H}}_{CS}^U} )^T}$, respectively, as
\begin{equation}\label{equ:Y_Uxr}
	{\bf{R}}_{{\bf{x}},r}^U{\rm{ = }}\frac{1}{{M_s^U}}{\bf{\hat H}}_{CS}^U{( {\bf{\hat H}}_{CS}^U )^H} \in \mathbb{C}^{N_c^U \times N_c^U},
\end{equation}
\begin{equation}\label{equ:Y_Uxf}
	{\bf{R}}_{{\bf{x}},f}^U{\rm{ = }}\frac{1}{{N_c^U}}{( {\bf{\hat H}}_{CS}^U )^T}{( {\bf{\hat H}}_{CS}^U )^{\rm{*}}} \in \mathbb{C}^{M_s^U \times M_s^U}.
\end{equation}
By applying eigenvalue decomposition to ${\bf{R}}_{{\bf{x}},r}^U$ and ${\bf{R}}_{{\bf{x}},f}^U$, we obtain 
\begin{equation}\label{equ:R_Uxr}
	\begin{aligned}
		[ {{\bf{U}}_{x,r}^U,{\bf{\Sigma }}_{x,r}^U} ] &= \text{eig}( {{\bf{R}}_{{\bf{x}},r}^U} ),\\
		[ {{\bf{U}}_{x,f}^U,{\bf{\Sigma }}_{x,f}^U} ] &= \text{eig}( {{\bf{R}}_{{\bf{x}},f}^U} ),
	\end{aligned}
\end{equation}
respectively, where ${\bf{\Sigma }}_{x,r}^U$ and ${\bf{\Sigma }}_{x,f}^U$ are the eigenvalue diagonal matrices, ${\bf{U}}_{x,r}^U$ and ${\bf{U}}_{x,f}^U$ are the corresponding eigen matrices, respectively. The number of targets is denoted by $N_{x,s}^U$. The noise subspaces of ${{\bf{R}}_{{\bf{x}},r}^U}$ and ${{\bf{R}}_{{\bf{x}},f}^U}$ are ${\bf{U}}_{x,rN}^U = {[ {{\bf{U}}_{x,r}^U} ]_{:,N_{x,s}^U + 1:N_c^U}}$ and ${\bf{U}}_{x,fN}^U = {[ {{\bf{U}}_{x,f}^U} ]_{:,N_{x,s}^U + 1:M_s^U}}$, respectively. 
The range and Doppler spectrum functions are, respectively, given by
\begin{equation}\label{equ:f_Ur}
	f_r^U\left( r \right) = {\bf{a}}_r^U{\left( r \right)^H}{\bf{U}}_{x,rN}^U{( {{\bf{U}}_{x,rN}^U} )^H}{\bf{a}}_r^U\left( r \right),
\end{equation}
\begin{equation}\label{equ:f_Uf}
	f_f^U\left( f \right) = {\bf{a}}_f^U{\left( f \right)^H}{\bf{U}}_{x,fN}^U{( {{\bf{U}}_{x,fN}^U} )^H}{\bf{a}}_f^U\left( f \right).
\end{equation}
where ${\bf{a}}_r^U\left( r \right) = {{\bf{a}}_r}( {r;\Delta {f^U},N_c^U} )$, and ${\bf{a}}_f^U\left( f \right) = {{\bf{a}}_f}( {f;T_s^U,M_s^U} )$. 
Furthermore, the range and Doppler spectra are, respectively, given by
\begin{equation}\label{equ:S_Ur}
	S_r^U\left( r \right) = {[ {{\bf{a}}_r^U{{\left( r \right)}^H}{\bf{U}}_{x,rN}^U{{({\bf{U}}_{x,rN}^U)}^H}{\bf{a}}_r^U\left( r \right)} ]^{ - 1}},
\end{equation}
\begin{equation}\label{equ:S_Uf}
	S_f^U\left( f \right) = {[ {{\bf{a}}_f^U{{\left( f \right)}^H}{\bf{U}}_{x,fN}^U{{({\bf{U}}_{x,fN}^U)}^H}{\bf{a}}_f^U\left( f \right)} ]^{ - 1}}.
\end{equation}

According to \textbf{Theorem~\ref{Theo:range_Doppler}}, the minimal points of $f_r^U\left( r \right)$ or $f_f^U\left( f \right)$, i.e., the maximal points of $S_r^U\left( r \right)$ or $S_f^U\left( f \right)$ are the range and Doppler estimation results.
\textbf{Algorithm~\ref{alg:Two-step_descent}} can be used to identify the above minimal values by reducing the second parameter, $\theta$, in \textbf{Algorithm~\ref{alg:Two-step_descent}} to be a {\color{blue} redundant} constant~\cite{Chen2023JCAS}. Note that $f\left( {{{\bf{p}}}} \right)$, ${\nabla _{\bf{p}}}f\left( {{{\bf{p}}}} \right)$ and ${{{\bf{H}}_{\bf{p}}}\left( {{{\bf{p}}}} \right)}$ in \textbf{Algorithm~\ref{alg:Two-step_descent}} are replaced by \eqref{equ:f_Ur}, $\frac{{\partial {f_r^U}\left( r \right)}}{{\partial r}}$, and $\frac{{{\partial ^2}{f_r^U}\left( r \right)}}{{{\partial ^2}r}}$ for range estimation, and replaced by \eqref{equ:f_Uf}, $\frac{{\partial {f_f^U}\left( f \right)}}{{\partial f}}$, and  $\frac{{{\partial ^2}{f_f^U}\left( f \right)}}{{{\partial ^2}f}}$ for Doppler estimation, respectively. 

The estimated range and Doppler sets are denoted by $\Theta _r^U = { {[ {\hat r_{k1}^U} ]} |_{k1 = 0, \cdots ,N_{x,s}^U - 1}}$ and $\Theta _f^U = { {[ {\hat f_{k2}^U} ]} |_{k2 = 0, \cdots ,N_{x,s}^U - 1}}$, respectively. Then, we provide the range-Doppler matching method to match the decoupled range and Doppler estimation results.

\subsubsection{Range-Doppler Matching Method}
The range and Doppler results are matched according to \textbf{Theorem~\ref{Theo:match}}.
\begin{Theo} \label{Theo:match}
	{\rm If ${\bf{\hat H}} = {{\bf{A}}_{\bf{r}}} {\bf{S}}{\left( {{{\bf{A}}_{\bf{f}}}} \right)^T} + {\bf{W}}$, where ${\bf{W}}$ is a Gaussian noise matrix, ${\bf{S}}$ is a diagonal matrix irrelevant to ${{\bf{A}}_{\bf{r}}}$ and ${{{\bf{A}}_{\bf{f}}}}$, ${{\bf{A}}_{\bf{r}}} = [{{\bf{a}}_r}({r_l};\Delta f,{N_c})]{|_{l = 0,1,...,L - 1}}$, and ${{\bf{A}}_{\bf{f}}} = [{{\bf{a}}_f}({f_l};{T_s},{M_s})]{|_{l = 0,1,...,L - 1}}$, then only the range-Doppler pair of the same target, denoted by $({r = r_l, f = f_l})$, {\color{blue} can achieve the maximal points of $\| {{{\left[ {{{\bf{a}}_r}\left( r \right)} \right]}^H}{{\bf{A}}_{\bf{r}}}{{\bf{S}}_S}{{\left( {{{\bf{A}}_{\bf{f}}}} \right)}^T}{{\left[ {{{\bf{a}}_f}\left( f \right)} \right]}^*}} \|_2^2$.}
		
	\begin{proof}			
		The proof is provided in \textbf{Appendix~\ref{appendix:match}}. 
	\end{proof}
	}

\end{Theo}

According to \textbf{Theorem~\ref{Theo:match}}, we define the range-Doppler matching matrix for UL JCAS as
\begin{equation}\label{equ:M_Utf}
	{\bf{M}}_{\tau f}^U{\rm{ = }}\| {{{[ {{\bf{\tilde A}}_r^U} ]}^H}{\bf{\hat H}}_{CS}^U{{[ {{\bf{\tilde A}}_f^U} ]}^*}} \|_2^2 \in \mathbb{C}^{N^U_{x,s} \times N^U_{x,s}},
\end{equation}
where ${\bf{\tilde A}}_r^U = { {\left[ {{{\bf{a}}_r}\left( {{{r}};\Delta {f^U},N_c^U} \right)} \right]} |_{r \in \Phi _r^U}}$ and ${\bf{\tilde A}}_f^U = { {[{{{\bf{a}}_f}\left( { f;T_s^U,M_s^U} \right)}  ]} |_{f \in \Phi _f^U}}$ are the estimated range and Doppler steering matrices for UL JCAS, respectively. The maximum of the $n$th row of ${\bf{M}}_{\tau f}^U$, e.g., ${[ {{\bf{M}}_{\tau f}^U} ]_{n,{m_n}}}$, indicates the $n$th value of $\Phi _r^U$ matches the ${m_n}$th value of $\Phi _f^U$.

\subsection{DLP CSI Estimation}\label{sec:DLP}
In the DLP period, the user receives the preamble signals from BS using receive BF vector, denoted by ${\bf{w}}_{RX}^D$. The received preamble signal is given by
\begin{equation}\label{equ:y_c_D}
	\begin{array}{l}
		\bar y_{C,n,m}^D = {( {{\bf{w}}_{RX}^D} )^H}{\bf{\bar y}}_{C,n,m}^D\\
		= \sqrt {\bar P_t^D} \bar d_{n,m}^D\sum\limits_{l = 0}^{L - 1} {\left[ \begin{array}{l}
				{b_{C,l}}{e^{j2\pi mT_s^D f_{c,d,l}}}\\
				\times {e^{ - j2\pi n\Delta {f^D} \tau _{c,l}}}\\
				\times \chi _{TX,l}^D\varpi _{RX,l}^D
			\end{array} \right]}  + n_{t,n,m}^D
	\end{array}
\end{equation}
where ${\bf{\bar y}}_{C,n,m}^D$ is given in \eqref{equ:y_C^U}, and $n_{t,n,m}^D{\rm{ = }}{( {{\bf{w}}_{RX}^D} )^H}{\bf{n}}_{t,n,m}^D$. Moreover, due to channel reciprocity, we have ${\bf{w}}_{RX}^D = {\bf{w}}_{TX}^U$ and ${\bf{w}}_{TX}^D = {\bf{w}}_{RX}^U$. Therefore, $\varpi _{RX,l}^D = \chi _{TX,l}^U$ and $\chi _{TX,l}^D = \varpi _{RX,l}^U$. According to \cite{2010MIMO}, the DL communication CSI obtained with the LS method is 
\begin{equation}\label{equ:hCS_D}
	\hat h_{CS,n,m}^D = \frac{{\bar y_{C,n,m}^D}}{{\sqrt {\bar P_t^D} \bar d_{n,m}^D}} = h_{CS,n,m}^D + w_{t,n,m}^D,
\end{equation}
where $h_{CS,n,m}^D = h_{CS,n,m}^U$ is due to the channel reciprocity, and $w_{t,n,m}^D = {{n_{t,n,m}^D} \mathord{/
		{\vphantom {{n_{t,n,m}^D} {(\sqrt {\bar P_t^D} \bar d_{n,m}^D)}}} 
		\kern-\nulldelimiterspace} {(\sqrt {\bar P_t^D} \bar d_{n,m}^D)}}$ is the transformed noise.



\subsection{DLD Period Mono-static JCAS Processing}
{\color{blue} In this subsection, we first present the received DL data and echo signals after BF, then propose the DLD JCAS BF and sensing schemes.}

\subsubsection{Received DL Data and Echo Signals}
In the DLD period, BS transmits data signals to the user using the communication link formed in the DLP period. Except for data communication, BS also transmits dedicated sensing beam to DoI, ${\bf{p}}_S^D$, with BF vector, ${{\bf{w}}_{TX}^{DS}}$. {\color{blue} The value of ${{\bf{w}}_{TX}^{DS}}$ should be optimized to minimize the interference to DL communication.}

The DL data signal received by the user is expressed as
\begin{equation}\label{equ:DLP_echo_match}
\begin{array}{l}
	y_{C,n,m}^D = {\bf{h}}_{C,n,m}^D \!\!\left[\!\! \begin{array}{l}
		d_{n,m}^D\sqrt {P_t^D} {\bf{w}}_{TX}^D\\
		+ d_{n,m}^{DS}\sqrt {P_t^{DS}} {\bf{w}}_{TX}^{DS}
	\end{array} \!\!\right] + w_{C,n,m}^D\\
	= d_{n,m}^D\sqrt {P_t^D} h_{CS,n,m}^D \!+\! d_{n,m}^{DS}\sqrt {P_t^{DS}} {\bf{h}}_{C,n,m}^D{\bf{w}}_{TX}^{DS} \!+\! w_{C,n,m}^D,
\end{array}
\end{equation}
where ${\bf{h}}_{C,n,m}^D = {( {{\bf{w}}_{RX}^D} )^H}{\bf{H}}_{C,n,m}^D \in \mathbb{C}^{1 \times {P_t}{Q_t}}$, $w_{C,n,m}^D = {( {{\bf{w}}_{RX}^D} )^H}{\bf{n}}_{t,n,m}^D$, and $d_{n,m}^{DS}$ and $P_t^{DS}$ are the transmit probe symbol and power, respectively. Note that the second term in \eqref{equ:DLP_echo_match} is the interference to DL communication.

In this process, ${{\bf{w}}_{TX}^{DS}}$ should be interference-free to DL communication. Therefore, ${{\bf{w}}_{TX}^{DS}}$ is in the nullspace of ${\bf{h}}_{C,n,m}^D$. Due to the channel reciprocity, we can use ${({\bf{\hat h}}_{C,n,m}^U)^T}$ to replace ${\bf{h}}_{C,n,m}^D$. By deriving the singular value decomposition of ${({\bf{\hat h}}_{C,n,m}^U)^T}$, the right singular matrix is obtained as ${{\bf{V}}_{C,n,m}^D}$, and the nullspace basis can be derived as 
\begin{equation}\label{equ:V_C_DN}
	{\bf{V}}_{C,n,m}^{DN} = {[ {{\bf{V}}_{C,n,m}^D} ]_{:,2:{P_t}{Q_t}}} \in \mathbb{C}^{{P_t}{Q_t} \times ( {{P_t}{Q_t} - 1} )}.
\end{equation}
Then, {\color{blue} ${\bf{w}}_{TX}^{DS}$ should be the linear transform of ${\bf{V}}_{C,n,m}^{DN}$ to guarantee no interference to the DL communication,} i.e., 
\begin{equation}\label{equ:W_TX_DS}
	{\bf{w}}_{TX}^{DS} = {\bf{V}}_{C,n,m}^{DN}{{\bf{m}}_1},
\end{equation} 
where ${\bf{m}}_1 \in \mathbb{C}^{( {{P_t}{Q_t} - 1} ) \times 1}$ and will be finally determined in the following subsection. 

On the other hand, to receive both the echo signals of the DL communication signal and those of the dedicated sensing signal, BS generates a BF matrix, ${\bf{W}}_{RX}^{DS} = [ {{\bf{w}}_{n,m}^D},{\bf{w}}_{n,m}^{DS}] \in \mathbb{C}^{{P_t}{Q_t} \times 2}$, to distinguish the echo signals from these two directions. The echo signals received at BS in the $n$th subcarrier of the $m$th OFDM symbol is expressed as ${\bf{r}}_{S,n,m}^{DS} = {( {{\bf{W}}_{RX}^{DS}} )^H}{\bf{y}}_{S,n,m}^{DS}$. Combining \eqref{equ:y_S^DS}, we express the sensing echo signal as
\begin{equation}\label{equ:r_S_nm}
{\bf{r}}_{S,n,m}^{DS} \!\! = \! {({\bf{W}}_{RX}^{DS})^H}{\bf{H}}_{S,n,m}^D \!\!\left( \!\!\! \begin{array}{l}
	\sqrt {P_t^D} d_{n,m}^D{\bf{w}}_{TX}^D +\\
	\sqrt {P_t^{DS}} d_{n,m}^{DS}{\bf{w}}_{TX}^{DS}
\end{array} \!\!\! \right) \!\!+\! {\bf{\bar n}}_{S,n,m}^{DS},
\end{equation}
where ${\bf{\bar n}}_{S,n,m}^{DS} = {( {{\bf{W}}_{RX}^{DS}} )^H}{\bf{n}}_{S,n,m}^{DS}$. {\color{blue} Note that ${\bf{w}}_{TX}^D$ generates the JCAS beam pointed at the estimated user's direction, $\hat{\bf{p}}_0^U$;} ${\bf{w}}_{TX}^{DS}$ generates the beam pointed at DoI, ${\bf{p}}_S^D$; and ${\bf{W}}_{RX}^{DS}$ is aimed to distinguish the echo signals from these two directions. 

\subsubsection{DLD JCAS BF Method}
Since ${\bf{H}}_{S,n,m}^D$ is unknown before transmitting JCAS signals, we have to use reference channel responses to generate ${\bf{w}}_{TX}^{DS}$ and ${\bf{W}}_{RX}^{DS}$. {\color{blue} The reference channel responses can be generated via \textbf{Theorem~\ref{Theo:Reference_channel_response}} based on the estimated UE's AoA and range.}

\begin{Theo} \label{Theo:Reference_channel_response}
	{\rm The actual mono-static echo sensing channel, denoted by ${{\bf{H}}_{SU}}$, has the same form as \eqref{equ:JCS sensing channel}, and is composed of $K$ scatterers with direction ${\Theta} = {\left. {\left\{ {{{\bf{p}}_{S,k}}} \right\}} \right|_{k = 0,...,K - 1}}$. Its DoI is ${{\bf{p}}_S}$, and the reference channel response can be defined as 
	\begin{equation}\label{equ:H_RS}
		{{\bf{H}}_{RS}}{\rm{ = }}\sqrt {{{{\lambda ^2}} \mathord{\left/
					{\vphantom {{{\lambda ^2}} {\left[ {{{\left( {4\pi } \right)}^3}{{\left( {{r_E}} \right)}^4}} \right]}}} \right.
					\kern-\nulldelimiterspace} {\left[ {{{\left( {4\pi } \right)}^3}{{\left( {{r_E}} \right)}^4}} \right]}}} {\bf{a}}\left( {{{\bf{p}}_S}} \right){\bf{a}}_{}^T\left( {{{\bf{p}}_S}} \right),
	\end{equation}
	where ${r_E}$ is the expected range of the target. The receive and transmit BF vectors generated from ${{\bf{H}}_{RS}}$, denoted, respectively, by ${{\bf{w}}_{RX}}$ and ${{\bf{w}}_{TX}}$, have the following properties: The value set $\left\{ {{{\bf{w}}_{RX}},{{\bf{w}}_{TX}}} \right\} = \left\{ {{\bf{w}}_{RX}^0,{\bf{w}}_{TX}^0} \right\}$ that satisfies $\| {{{( {{\bf{w}}_{RX}^0} )}^H}{\bf{H}}_{RS}{\bf{w}}_{TX}^0} \|_2^2 = 0$ can make $\| {{{( {{\bf{w}}_{RX}^0} )}^H}{\bf{H}}_{SU}{\bf{w}}_{TX}^0} \|_2^2 \approx 0$; the value set $\left\{ {{{\bf{w}}_{RX}},{{\bf{w}}_{TX}}} \right\} = \{ {{\bf{w}}_{RX}^{max},{\bf{w}}_{TX}^{max}} \}$ that maximizes $\| {{{( {{{\bf{w}}_{RX}}} )}^H}{\bf{H}}_{RS}{{\bf{w}}_{TX}}} \|_2^2$ can also maximize $\| {{{( {{\bf{w}}_{RX}} )}^H}{\bf{H}}_{SU}{\bf{w}}_{TX}} \|_2^2$. 
		\begin{proof}			
			The proof is provided in \textbf{Appendix~\ref{appendix:Reference_channel}}. 
		\end{proof}
	}
\end{Theo}

According to \textbf{Theorem~\ref{Theo:Reference_channel_response}}, we can define the reference channel matrix for generating ${\bf{w}}_{n,m}^{DS}$ and ${\bf{w}}_{TX}^{DS}$ as 
\begin{equation}\label{equ:H_RS_D}
	{\bf{H}}_{RS,n,m}^D = \sqrt {{{{\lambda ^2}} \mathord{\left/
				{\vphantom {{{\lambda ^2}} {\left[ {{{(4\pi )}^3}{{({r_E})}^4}} \right]}}} \right.
				\kern-\nulldelimiterspace} {\left[ {{{(4\pi )}^3}{{({r_E})}^4}} \right]}}} {\bf{a}}\left( {{\bf{p}}_S^D} \right){\bf{a}}_{}^T\left( {{\bf{p}}_S^D} \right),
\end{equation}
where $r_E$ is the expected range of the target. Similarly, the reference channel matrix for generating ${\bf{w}}_{n,m}^D$ is defined as
\begin{equation}\label{equ:H_IS_nm}
	{\bf{H}}_{IS,n,m}^D = \sqrt {{{{\lambda ^2}} \mathord{\left/
				{\vphantom {{{\lambda ^2}} {\left[ {{{(4\pi )}^3}{{(\hat r_0^U)}^4}} \right]}}} \right.
				\kern-\nulldelimiterspace} {\left[ {{{(4\pi )}^3}{{(\hat r_0^U)}^4}} \right]}}} {\bf{a}}\left( {{\bf{\hat p}}_0^U} \right){\bf{a}}_{}^T\left( {{\bf{\hat p}}_0^U} \right),
\end{equation}
where ${\bf{\hat p}}_0^U$ and $\hat r_0^U$ are the estimated direction and range of the user estimated in Section~\ref{sec:UplinkJCAS}. Here, we set $r_E = \hat r_0^U$ to balance the propagation loss of two echo signals. The reference received signal can be expressed with ${\bf{H}}_{RS,n,m}^D$ and ${\bf{H}}_{IS,n,m}^D$, and is given by
\begin{equation}\label{equ:r_RS}
	{\bf{r}}_{RS,n,m}^{DS} = {({\bf{W}}_{RX}^{DS})^H}\left( \begin{array}{l}
		{\bf{H}}_{IS,n,m}^D\sqrt {P_t^D} d_{n,m}^D{\bf{w}}_{TX}^D + \\
		{\bf{H}}_{RS,n,m}^D\sqrt {P_t^{DS}} d_{n,m}^{DS}{\bf{w}}_{TX}^{DS}
	\end{array} \right).
\end{equation}
Since ${\bf{w}}_{TX}^{DS} = {\bf{V}}_{C,n,m}^{DN}{{\bf{m}}_1}$ and ${\bf{W}}_{RX}^{DS} = [ {{\bf{w}}_{n,m}^D,{\bf{w}}_{n,m}^{DS}} ]$ are designed by maximizing the received signals while eliminating the interference, the criterion to generate ${\bf{w}}_{TX}^{DS}$ and ${\bf{W}}_{RX}^{DS} = [ {{\bf{w}}_{n,m}^D,{\bf{w}}_{n,m}^{DS}} ]$ can be, respectively, given by
\begin{equation}\label{equ:P1}
	\begin{array}{l}
		\mathop {\max }\limits_{{\bf{w}}_{n,m}^D} \| {{{({\bf{w}}_{n,m}^D)}^H}{\bf{H}}_{IS,n,m}^D{\bf{w}}_{TX}^D} \|_2^2\\
		s.t.~{({\bf{w}}_{n,m}^D)^H}{\bf{H}}_{RS,n,m}^D = {\bf{0}}\\
		\quad\quad \left\| {{\bf{w}}_{n,m}^D} \right\|_2^2 = 1
	\end{array},
\end{equation}
\begin{equation}\label{equ:P2}
	\begin{array}{l}
		\mathop {\max }\limits_{{\bf{w}}_{n,m}^{DS},{{\bf{m}}_1}} \| {{{({\bf{w}}_{n,m}^{DS})}^H}{\bf{H}}_{RS,n,m}^D{\bf{V}}_{C,n,m}^{DN}{{\bf{m}}_1}} \|_2^2\\
		\quad\quad\quad s.t.~{({\bf{w}}_{n,m}^{DS})^H}{\bf{H}}_{IS,n,m}^D = {\bf{0}}\\
		\quad\quad\quad \left\| {{\bf{w}}_{n,m}^{DS}} \right\|_2^2 = \left\| {{{\bf{m}}_1}} \right\|_2^2 = 1
	\end{array}.
\end{equation}

{\color{blue} 
	The derivation of ${\bf{w}}_{n,m}^D$, ${\bf{w}}_{n,m}^{DS}$, and ${\bf{w}}_{TX}^{DS}$ is provided in \textbf{Appendix~\ref{appendix:derivation_P1P2}}. The final solutions to them are 
	\begin{equation}\label{equ:w_nm_D_DS}
		\begin{aligned}
			{\bf{w}}_{n,m}^D &= {\bf{U}}_{RS,n,m}^{DN}{\left[ {{\bf{U}}_{IS}^D} \right]_{:,1}},\;\\
			{\bf{w}}_{n,m}^{DS} &= {\bf{U}}_{IS,n,m}^{DN}{\left[ {{\bf{U}}_{RS}^D} \right]_{:,1}}, \;\\
			{\bf{w}}_{TX}^{DS} &= {\bf{V}}_{C,n,m}^{DN}{\left[ {{\bf{V}}_{RS}^D} \right]_{:,1}},
		\end{aligned}
	\end{equation}
	where ${\bf{U}}_{RS,n,m}^{DN}$ and ${\bf{U}}_{IS,n,m}^{DN}$ are the left singular matrices of ${\bf{H}}_{RS,n,m}^D$ and ${\bf{H}}_{IS,n,m}^D$, respectively; ${\bf{U}}_{IS}^D$ is the left singular matrix of ${( {{\bf{U}}_{RS,n,m}^{DN}} )^H}{\bf{H}}_{IS,n,m}^D{\bf{w}}_{TX}^D$, and ${\bf{U}}_{RS}^D$ and ${{\bf{V}}_{RS}^D}$ are the left and right singular matrices of ${({\bf{U}}_{IS,n,m}^{DN})^H}{\bf{H}}_{RS,n,m}^D{\bf{V}}_{C,n,m}^{DN}$.
}

Substituting \eqref{equ:w_nm_D_DS} into \eqref{equ:r_S_nm} and \eqref{equ:DLP_echo_match}, the received DL communication signals and echo signals are formed completely. 
Subsequently, we present the DLD period sensing signal processing methods.

\subsubsection{DLD Period Sensing Signal Processing}
Since \eqref{equ:w_nm_D_DS} are solutions to the problems \eqref{equ:P1} and \eqref{equ:P2}, according to \textbf{Theorem \ref{Theo:Reference_channel_response}}, ${[ {{\bf{r}}_{S,n,m}^{DS}} ]_1}$ and ${[ {{\bf{r}}_{S,n,m}^{DS}} ]_2}$ are the echo signals from directions, ${\bf{\hat p}}_0^U$ and ${\bf{p}}_S^D$, respectively. In order to obtain the ranges and Doppler frequencies of targets from ${ {{\bf{r}}_{S,n,m}^{DS}}}$, the transmit symbols are removed first, and we obtain $\hat h_{S1,n,m}^{DS} = {{{{[ {{\bf{r}}_{S,n,m}^{DS}} ]}_1}} \mathord{/
		{\vphantom {{{{[ {{\bf{r}}_{S,n,m}^{DS}} ]}_1}} {d_{n,m}^D}}} 
		\kern-\nulldelimiterspace} {d_{n,m}^D}}$ and $\hat h_{S2,n,m}^{DS} = {{{{[ {{\bf{r}}_{S,n,m}^{DS}} ]}_2}} \mathord{/
		{\vphantom {{{{[ {{\bf{r}}_{S,n,m}^{DS}} ]}_2}} {d_{n,m}^{DS}}}} 
		\kern-\nulldelimiterspace} {d_{n,m}^{DS}}}$. According to \eqref{equ:r_S_nm}, we have
\begin{equation}\label{equ:h_S1_nm}
		\hat h_{S1,n,m}^{DS} \!\!= \!\!\sum\limits_{l = 0}^{L - 1} \!{\left[\!\! \begin{array}{l}
				{b_{S,l}}\chi _{RX,l}^D\chi _{TX,l}^D\sqrt {P_t^D} \\
				\times {e^{j2\pi {f_{s,l,1}}mT_s^D}}{e^{ - j2\pi n\Delta {f^D} {{\tau _{s,l}}} }}
			\end{array} \!\! \right]}  \!\! + \!\! N_{S1,n,m}^{DS},
\end{equation}
\begin{equation}\label{equ:h_S2_nm}
	\hat h_{S2,n,m}^{DS} \!\!=\!\! \sum\limits_{k = 0}^{K - 1}\!\! {\left[ \begin{array}{l}
			{b_{S,k}}\chi _{RX,k}^{DS}\chi _{TX,k}^{DS}\sqrt {P_t^{DS}} \\
			\times {e^{j2\pi mT_s^D {f_{d,k}^{SU}} }}{e^{ - j2\pi n\Delta {f^D} {\frac{{2r_k^{RS}}}{c}} }}
		\end{array} \right]}  + N_{S2,n,m}^{DS},
\end{equation}
respectively, where $K$ is the number of targets in DoI, ${f_{d,k}^{SU}}$ and ${r_k^{RS}}$ are the Doppler and range of the $k$th target in DoI; $\chi _{RX,l}^D = {( {{\bf{w}}_{n,m}^D} )^H}{\bf{a}}( {{\bf{p}}_{RX,l}^{DS}} )$, $\chi _{TX,l}^D = {{\bf{a}}^T}( {{\bf{p}}_{TX,l}^D} ){\bf{w}}_{TX}^D$, $\chi _{RX,k}^{DS} = {( {{\bf{w}}_{n,m}^{DS}} )^H}{\bf{a}}( {{\bf{p}}_{S,k}^D} )$, and $\chi _{TX,k}^{DS} = {\bf{a}}_{}^T( {{\bf{p}}_{S,k}^D} ){\bf{w}}_{TX}^{DS}$ are the BF gains. Moreover,
$N_{S1,n,m}^{DS} = {{{{( {{\bf{w}}_{n,m}^D} )}^H}( {{\bf{n}}_{S,n,m}^{DS}} )} \mathord{/
		{\vphantom {{{{( {{\bf{w}}_{n,m}^D} )}^H}( {{\bf{n}}_{S,n,m}^{DS}} )} {d_{n,m}^D}}} 
		\kern-\nulldelimiterspace} {d_{n,m}^D}}$ and $N_{S2,n,m}^{DS} = {{{{( {{\bf{w}}_{n,m}^{DS}} )}^H}( {{\bf{n}}_{S,n,m}^{DS}} )} \mathord{/
		{\vphantom {{{{( {{\bf{w}}_{n,m}^{DS}} )}^H}( {{\bf{n}}_{S,n,m}^{DS}} )} {d_{n,m}^{DS}}}} 
		\kern-\nulldelimiterspace} {d_{n,m}^{DS}}}$ are the equivalent noises.

After $M_s^D$ OFDM symbols at $N_c^D$ subcarriers are transmitted, we obtain echo signal matrices ${\bf{\hat H}}_{S1}^{DS}$ and ${\bf{\hat H}}_{S2}^{DS}$, where ${[ {{\bf{\hat H}}_{S1}^{DS}} ]_{n,m}} = \hat h_{S1,n,m}^{DS}$ and ${[ {{\bf{\hat H}}_{S2}^{DS}} ]_{n,m}} = \hat h_{S2,n,m}^{DS}$. We can see that ${\bf{\hat H}}_{S1}^{DS}$ and ${\bf{\hat H}}_{S2}^{DS}$ are also composed of range and Doppler steering vectors as shown in \eqref{equ:aUr} and \eqref{equ:aUf}. Here, we construct range and Doppler steering vector for DL echo sensing as
\begin{equation}\label{equ:a_r^D}
	\begin{aligned}
		{\bf{a}}_r^D\left( r \right) &= {{\bf{a}}_r}\left( {r;\Delta {f^D},N_c^D} \right),\\
		{\bf{a}}_f^D\left( f \right) &= {{\bf{a}}_f}\left( {f;T_s^D,M_s^D} \right).
	\end{aligned}
\end{equation}
According to \textbf{Theorem~\ref{Theo:range_Doppler}}, we use the noise subspaces of $\frac{1}{{M_s^D}}{\bf{\hat H}}_{S1}^{DS}{( {{\bf{\hat H}}_{S1}^{DS}} )^H}$ and $\frac{1}{{M_s^D}}{\bf{\hat H}}_{S2}^{DS}{( {{\bf{\hat H}}_{S2}^{DS}} )^H}$, denoted by ${\bf{U}}_{x,rN}^{DS1}$ and ${\bf{U}}_{x,rN}^{DS2}$, to construct the range spectrum functions as
\begin{equation}\label{equ:f_r_DS}
	\begin{aligned}
		f_{r1}^{DS}\left( r \right) = {\left[ {{\bf{a}}_r^D\left( r \right)} \right]^H}{\bf{U}}_{x,rN}^{DS1}{\left( {{\bf{U}}_{x,rN}^{DS1}} \right)^H}{\bf{a}}_r^D\left( r \right),\\
		f_{r2}^{DS}\left( r \right) = {\left[ {{\bf{a}}_r^D\left( r \right)} \right]^H}{\bf{U}}_{x,rN}^{DS2}{\left( {{\bf{U}}_{x,rN}^{DS2}} \right)^H}{\bf{a}}_r^D\left( r \right).	\end{aligned}
\end{equation}
We also use the noise subspaces of $\frac{1}{{N_c^D}}{( {{\bf{\hat H}}_{S1}^{DS}} )^T}{( {{\bf{\hat H}}_{S1}^{DS}} )^*}$ and $\frac{1}{{N_c^D}}{( {{\bf{\hat H}}_{S2}^{DS}} )^T}{( {{\bf{\hat H}}_{S2}^{DS}} )^*}$, denoted by ${\bf{U}}_{x,fN}^{DS1}$ and ${\bf{U}}_{x,fN}^{DS2}$, to construct the Doppler spectrum functions as
\begin{equation}\label{equ:f_f_DS}
	\begin{aligned}
		f_{f1}^{DS}\left( f \right) = {\left[ {{\bf{a}}_f^D\left( f \right)} \right]^H}{\bf{U}}_{x,fN}^{DS1}{\left( {{\bf{U}}_{x,fN}^{DS1}} \right)^H}{\bf{a}}_f^D\left( f \right),\\
		f_{f2}^{DS}\left( f \right) = {\left[ {{\bf{a}}_f^D\left( f \right)} \right]^H}{\bf{U}}_{x,fN}^{DS2}{\left( {{\bf{U}}_{x,fN}^{DS2}} \right)^H}{\bf{a}}_f^D\left( f \right).
	\end{aligned}
\end{equation}
Then, we can use \textbf{Algorithm~\ref{alg:Two-step_descent}} to identify the minimal points of $f_{r1}^{DS}\left( r \right)$ and $f_{f1}^{DS}\left( f \right)$, and $f_{r2}^{DS}\left( r \right)$ and $f_{f2}^{DS}\left( f \right)$ as processed in Section~\ref{sec:UL_JCAS_range_Doppler}. These minimal points are the estimated range and Doppler results for the targets in directions ${\bf{\hat p}}_0^U$ and ${\bf{p}}_S^D$, respectively. Note that ${\nabla _{\bf{p}}}f\left( {{{\bf{p}}}} \right)$ and ${{{\bf{H}}_{\bf{p}}}\left( {{{\bf{p}}}} \right)}$ in \textbf{Algorithm~\ref{alg:Two-step_descent}} are replaced with the inverse value, the first-order and second-order derivatives of $f_{r1}^{DS}\left( r \right)$ and $f_{r2}^{DS}\left( r \right)$ for range estimation, and those of $f_{f1}^{DS}\left( f \right)$ and $f_{f2}^{DS}\left( f \right)$ for Doppler estimation, respectively. 

The range estimation is denoted by $\Theta _{r1}^{DS} = { {[ {{{\hat r_{k1}^{DS}} \mathord{/
					{\vphantom {{\hat r_{k1}^{DS}} 2}} 
					\kern-\nulldelimiterspace} 2}} ]} |_{k1 = 0, \cdots ,N_{x,s1}^D - 1}}$, and the Doppler estimation is denoted by $\Theta _{f1}^{DS} = { {[ {{{\hat f_{k1}^{DS}} \mathord{/
					{\vphantom {{\hat f_{k1}^{DS}} 2}} 
					\kern-\nulldelimiterspace} 2}} ]} |_{k1 = 0, \cdots ,N_{x,s1}^D - 1}}$. The range and Doppler estimation sets of the targets in direction ${\bf{p}}_S^D$ are denoted by $\Theta _{r2}^{DS} = { {[ {{{\hat r_{k2}^{DS}} \mathord{/
					{\vphantom {{\hat r_{k2}^{DS}} 2}} 
					\kern-\nulldelimiterspace} 2}} ]} |_{k2 = 0, \cdots ,N_{x,s2}^D - 1}}$ and $\Theta _{f2}^{DS} = { {[ {{{\hat f_{k2}^{DS}} \mathord{/
					{\vphantom {{\hat f_{k2}^{DS}} 2}} 
					\kern-\nulldelimiterspace} 2}} ]} |_{k2 = 0, \cdots ,N_{x,s2}^D - 1}}$, respectively, where $N_{x,s1}^D$ and $N_{x,s2}^D$ are the numbers of targets in the corresponding directions, respectively. 
Then, we match the decoupled range and Doppler estimation results according to~\textbf{Theorem~\ref{Theo:match}}. 
The matching matrices can be constructed as
\begin{equation}\label{equ:M_rf}
	\begin{aligned}
		{\bf{M}}_{rf1}^{DS}{\rm{ = }}\| {{{[ {{\bf{\tilde A}}_{r1}^{DS}} ]}^H}{\bf{\hat H}}_{S1}^{DS}{{[ {{\bf{\tilde A}}_{f1}^{DS}} ]}^*}} \|_2^2 \in \mathbb{C}^{N_{x,s1}^D \times N_{x,s1}^D}\\
		{\bf{M}}_{rf2}^{DS}{\rm{ = }}\| {{{[ {{\bf{\tilde A}}_{r2}^{DS}} ]}^H}{\bf{\hat H}}_{S2}^{DS}{{[ {{\bf{\tilde A}}_{f2}^{DS}} ]}^*}} \|_2^2 \in \mathbb{C}^{N_{x,s2}^D \times N_{x,s2}^D}
	\end{aligned},
\end{equation}
where ${\bf{\tilde A}}_{r1}^{DS} = {{[ {{\bf{a}}_r^D\left( {2r} \right)} ]} |_{r1 \in \Theta _{r1}^{DS}}}$, ${\bf{\tilde A}}_{f1}^{DS} = { {[ {{\bf{a}}_f^D\left( {2f} \right)} ]} |_{f \in \Theta _{f1}^{DS}}}$, ${\bf{\tilde A}}_{r2}^{DS} = { {[ {{\bf{a}}_r^D\left( {2r} \right)} ]} |_{r2 \in \Theta _{r2}^{DS}}}$, and ${\bf{\tilde A}}_{f2}^{DS} = { {[ {{\bf{a}}_f^D\left( {2f} \right)} ]} |_{f \in \Theta _{f2}^{DS}}}$.

The maximal value of the $n$th row of ${\bf{M}}_{rf}$ (${\bf{M}}_{rf}$ can be ${\bf{M}}_{rf1}^{DS}$ or ${\bf{M}}_{rf2}^{DS}$), e.g., ${[ {{\bf{M}}_{\tau f}} ]_{n,{m_n}}}$, indicates the $n$th point in the range set matches the ${m_n}$th point in the Doppler set. 
Next, $\Theta _{r1}^{DS}$, $\Theta _{f1}^{DS}$, $\Theta _{r2}^{DS}$, and $\Theta _{f2}^{DS}$ are rearranged by the matching result.
		
\section{DUC JCAS Data Fusion Method} \label{sec:DUC_JCAS_Data_Fusion}
{\color{blue} In the consecutive UL and DL time slots, due to the channel reciprocity, the relative location and Doppler between BS and the targets, and the UL and DL CSI are treated as unchanged in the block. This section presents the DUC JCAS fusion method for sensing data fusion and communication CSI refining based on this feature.}

\subsection{Estimation Feature Acquisition for DUC JCAS}

The location of a detected target can be derived, respectively, as
\begin{equation}\label{equ:p_loc}
		{\bf{\Omega }} = {( {r\sin \theta \cos \varphi ,r\sin \theta \sin \varphi ,r\cos \theta } )^T},
\end{equation}
where $r$ and ${\bf{p}} = \left( {\theta ,\varphi } \right)$ are the estimated range and direction, respectively. 

In the ULP period, since the LoS path dominates the UL JCAS channel, we only estimate the range and AoA of the user, i.e., $N_{x,s}^{U} = 1$. The range and AoA of the user are obtained as $\Theta _r^U = {\left. {[ {\hat r_k^U} ]} \right|_{k = 0}}$ and ${\bf{\hat p}}_0^U$, respectively, and the location of user is calculated as ${{\bf{\hat \Omega }}_0^U}$ as shown in~\eqref{equ:p_loc}. In the DLD period, the range and AoA results of the targets in DoU are $\Theta _{r1}^{DS} = {{[ {\frac{{\hat r_{k1}^{DS}}}{2}} ]} |_{k_1 = 0, \cdots ,N_{x,s1}^D - 1}}$ and ${\bf{\hat p}}_0^U$, respectively, and we calculate the location of the $k_1$th target as ${{\bf{\hat \Omega }}_{k1}^{DS}}$; the range and AoA results of the targets in DoI are $\Theta _{r2}^{DS} = {{[ {\frac{{\hat r_{k2}^{DS}}}{2}} ]} |_{k_2 = 0, \cdots ,N_{x,s2}^D - 1}}$ and ${\bf{p}}_S^D$, and the location of the $k_2$th target is calculated as ${{\bf{\hat \Omega }}_{k2}^{DS}}$. 

Choose the location and Doppler as the feature set for the sensing targets. The feature set for ULP JCAS targets is $\Phi _{}^U = {{\{ {{\bf{\hat \Omega }}_k^U,\hat f_k^U} \}} |_{k = 0}}$, the feature set for ULD JCAS targets in DoU is $\Phi _1^{DS} = { {\{ {{\bf{\hat \Omega }}_{k1}^{DS},\hat f_{k1}^{DS}} \}} |_{k1 = 0, \cdots ,N_{x,s1}^D - 1}}$, and the feature set for ULD JCAS targets in DoI is $\Phi _2^{DS} = { {\{ {{\bf{\hat \Omega }}_{k2}^{DS},\hat f_{k2}^{DS}} \}} |_{k2 = 0, \cdots ,N_{x,s2}^D - 1}}$.

\subsection{DUC JCAS Sensing Data Fusion Method} \label{sec:DUC_JCAS_SenDataFusion}
Notice that $\Phi^U$ is the sensing result of the user, while $\Phi _1^{DS}$ is the sensing results, including the user and other dumb targets. Therefore, we can distinguish between the user and other targets by comparing the points in $\Phi_1^{DS}$ and $\Phi^U$.

We first give a normalized distance measurement between two estimation feature sets, denoted by ${\Phi _1} = { {\{ {{{\bf{\Omega }}_{k1}},{f_{k1}}} \}} |_{k1 = 1, \cdots ,K1}}$ and ${\Phi _2} = { {\{ {{{\bf{\Omega }}_{k2}},{f_{k2}}} \}} |_{k2 = 1, \cdots ,K2}}$. The location and Doppler Euclidean distance matrices between ${\Phi _1}$ and ${\Phi _2}$ are given as ${{\bf{Z}}_{loc}} \in \mathbb{R}^{K1 \times K2}$ and ${{\bf{Z}}_f} \in \mathbb{R}^{K1 \times K2}$, respectively, with ${\left[ {{{\bf{Z}}_{loc}}} \right]_{k1,k2}} = \left\| {{{\bf{\Omega }}_{k1}} - {{\bf{\Omega }}_{k2}}} \right\|_2^2$ and ${\left[ {{{\bf{Z}}_f}} \right]_{k1,k2}} = \left\| {{f_{k1}} - {f_{k2}}} \right\|_2^2$. Then, we construct the normalized distance matrix as 
\begin{equation}\label{equ:Z}
	{\bf{Z}} = {{{{\bf{Z}}_{loc}}} \mathord{\left/
			{\vphantom {{{{\bf{Z}}_{loc}}} {{{\left[ {{{\bf{Z}}_{loc}}} \right]}_{\max }}}}} \right.
			\kern-\nulldelimiterspace} {{{\left[ {{{\bf{Z}}_{loc}}} \right]}_{\max }}}} + {{{{\bf{Z}}_f}} \mathord{\left/
			{\vphantom {{{{\bf{Z}}_f}} {{{\left[ {{{\bf{Z}}_f}} \right]}_{\max }}}}} \right.
			\kern-\nulldelimiterspace} {{{\left[ {{{\bf{Z}}_f}} \right]}_{\max }}}},
\end{equation}
where ${{{\left[ {{{\bf{Z}}_{loc}}} \right]}_{\max }}}$ and ${{{\left[ {{{\bf{Z}}_f}} \right]}_{\max }}}$ are the maximum values of ${{{\bf{Z}}_{loc}}}$ and ${{{\bf{Z}}_{f}}}$, respectively; and $[{\bf{Z}}]_{i,j} \in [0,2]$  for all the elements of ${\bf{Z}}$.

In the high signal-to-noise ratio (SNR) regime, since the estimation mean square error (MSE) of a target is much smaller than the square distance between two different targets, according to the maximum likelihood (ML) criterion, the point-pair between $\Phi_1$ and $\Phi_2$ with the least distance shall be matched as the same target~\cite{levy2008principles}. Therefore, if the minimum value of the $k$th row of ${\bf{Z}}$ is ${\left[ {\bf{Z}} \right]_{k,l}}$, then the $k$th point of ${\Phi _1}$ matches the $l$th point of ${\Phi _2}$. 

Replace $\Phi_1$ and $\Phi_2$ with $\Phi^U$ and $\Phi_1^{DS}$, and then the matched point is the user, while the rest are the dumb points. The matched points in $\Phi^U$ and $\Phi_1^{DS}$ can be treated as independent estimates of the same parameters. Therefore, the sensing results of a matched point pair can be fused to generate more accurate sensing results. The sensing data fusion method can be developed based on \textbf{Theorem~\ref{Theo:Fusion_criterion}}.

\begin{Theo} \label{Theo:Fusion_criterion}
	{\rm Two independent estimates of the same target are denoted by ${{\bf{v}}_1} = {\bf{v}} + \Delta {{\bf{v}}_1}$ and ${{\bf{v}}_2} = {\bf{v}} + \Delta {{\bf{v}}_2}$, respectively, where ${\bf{v}}$ is the actual value, $\Delta {{\bf{v}}_1}$ and $\Delta {{\bf{v}}_2}$ are errors that follow Gaussian distributions \cite{richards2010principles} with ${\bf{0}}$-mean and variance $E\{ \left\| {\Delta {{\bf{v}}_1}} \right\|_2^2\}  = {\sigma _1^2}$ and $E\{ \left\| {\Delta {{\bf{v}}_2}} \right\|_2^2\} = {\sigma _2^2}$, respectively. The fusion sensing result is ${\bf{\bar v}} = {{\bf{v}}_1} + \alpha ({{\bf{v}}_2} - {{\bf{v}}_1})$ ($0 < \alpha < 1$). Then, the optimal $\alpha$, denoted by $\alpha^*$, that minimizes $E\{ \left\| {{\bf{\bar v}}} \right\|_2^2\} $ is ${ \alpha^*} = \frac{{{\sigma _1^2}}}{{{\sigma _1^2} + {\sigma _2^2}}}$. The minimum of $E\{ \left\| {{\bf{\bar v}}} \right\|_2^2\} $ is $\frac{{{\sigma _1^2}{\sigma _2^2}}}{{{\sigma _1^2} + {\sigma _2^2}}}$.
		\begin{proof}			
			The proof is provided in \textbf{Appendix~\ref{appendix:FUSION_criterion}}. 
		\end{proof}
	}
\end{Theo}

{\color{blue} 
According to \textbf{Theorem~\ref{Theo:Fusion_criterion}}, we can merge the sensing results, including range and velocity, based on the estimation MSE of them. In practical applications, the estimation MSE is not easy to derive directly, we alternatively use Cramer-Rao lower bound (CRLB), i.e., the lower bound for estimation MSE. Moreover, the sensing CRLB is typically inversely proportional to the sensing SNR in the high SNR regime~\cite{CRLB2017}, hence we can use the inverse of sensing SNR to replace sensing CRLB to form a weighted sum of the sensing results in \textbf{Theorem~\ref{Theo:Fusion_criterion}}. 

The sensing SNR can be derived using the eigenvalues of ${\bf{\hat H}}_{CS}^U( {\bf{\hat H}}_{CS}^U )^H$ and ${\bf{\hat H}}_{S1}^{DS}{( {{\bf{\hat H}}_{S1}^{DS}} )^H}$~\cite{Chen2023JCAS}. The eigenvalue matrices of ${\bf{\hat H}}_{CS}^U( {\bf{\hat H}}_{CS}^U )^H$ and ${\bf{\hat H}}_{S1}^{DS}{( {{\bf{\hat H}}_{S1}^{DS}} )^H}$ in descending order are derived as ${\bf{\Sigma }}_r^U \in \mathbb{R}^{N_c^U \times 1}$ and ${\bf{\Sigma }}_{r1}^{DS} \in \mathbb{R}^{N_c^D \times 1}$, respectively.
Here, we derive the sensing SNR of $\Phi^U$ as an example. The sensing SNR of the $k$th point of $\Phi^U$ is 
\begin{equation}\label{equ:SNR_k}
	{\gamma _k} = {{({{[{\bf{\Sigma }}_r^U]}_k} - \hat \sigma _N^U)} \mathord{\left/
			{\vphantom {{({{[{\bf{\Sigma }}_r^U]}_k} - \hat \sigma _N^U)} {\hat \sigma _N^U}}} \right.
			\kern-\nulldelimiterspace} {\hat \sigma _N^U}},
\end{equation}
where $\hat \sigma _N^U$ is the estimated noise power and is calculated as the mean value of the last $N_c^U - N_{x,s}^U$ eigenvalues of ${\bf{\Sigma }}_r^U$.
In this way, all the sensing SNRs of detection points in $\Phi^U$ and $\Phi_1^{DS}$ can be estimated, and the inverse of sensing SNRs can be used as the variances in \textbf{Theorem~\ref{Theo:Fusion_criterion}}.
}

According to \textbf{Theorem~\ref{Theo:Fusion_criterion}}, we summarize the matching and fusion method in \textbf{Algorithm~\ref{DUC_JCAS_fusion_algorithm}}, and the output fused estimation set is $\bar \Phi $. Note that the first point in $\bar \Phi $ is the user. Finally, $\bar \Phi $ and $\Phi _2^{DS}$ are obtained as the sensing results in a round of consecutive UL and DL time slots.

\subsection{DUC Communication CSI Fusion Method}

{\color{blue} BS stores both the UL CSI estimates and DL CSI feedback. Here, we consider the feedback with no quantization error. } 
According to \eqref{equ:y_CS_nm} and \eqref{equ:hCS_D}, we can see that the UL and DL estimated CSI can also be treated as independent observation of the same CSI due to the channel reciprocity. 
Therefore, \textbf{Theorem~\ref{Theo:Fusion_criterion}} can also be used to refine the estimated CSI. Since $\left\| {{\bf{w}}_{RX}^D} \right\|_2^2 = \left\| {{\bf{w}}_{RX}^U} \right\|_2^2 = 1$, and $\bar d_{n,m}^D$ and $\bar d_{n,m}^U$ are the preamble symbols with constant modulus 1, we obtain the variance of $w_{t,n,m}^U$ and $w_{t,n,m}^D$ as $\sigma _1^2 = \frac{{\sigma _N^2}}{{\bar P_t^U}} = \frac{1}{{{\gamma _U}}}$ and $\sigma _2^2 = \frac{{\sigma _N^2}}{{\bar P_t^D}} = \frac{1}{{{\gamma _D}}}$, respectively, where ${\gamma _U}$ and ${\gamma _D}$ are the SNRs for the ULP and DLP communication received signals, respectively.

Construct ${\bf{\hat H}}_{CS}^U \in {\mathbb{C}^{N_c^U \times M_s^U}}$ and ${\bf{\hat H}}_{CS}^D \in {\mathbb{C}^{N_c^D \times M_s^D}}$, where ${[{\bf{\hat H}}_{CS}^U]_{n,m}} = \hat h_{CS,n,m}^U$ and ${[{\bf{\hat H}}_{CS}^D]_{n,m}} = \hat h_{CS,n,m}^D$. Based on ${\bf{\hat H}}_{CS}^U$ and ${\bf{\hat H}}_{CS}^D$, we can use the same SNR estimation method in \eqref{equ:SNR_k} to {\color{blue} calculate ${\gamma _U}$ and ${\gamma _D}$ from the eigenvalues of ${\bf{\hat H}}_{CS}^U{({\bf{\hat H}}_{CS}^U)^H}$ and ${\bf{\hat H}}_{CS}^D{({\bf{\hat H}}_{CS}^D)^H}$, respectively.} According to \textbf{Theorem~\ref{Theo:Fusion_criterion}}, we can fuse the communication CSI based on $\sigma _1^2 = \frac{1}{{{\gamma _U}}}$ and $\sigma _2^2 = \frac{1}{{{\gamma _D}}}$ as 
\begin{equation}\label{equ:SNR_fuse}
	\hat h_{CS,n,m}^{DU} = \hat h_{CS,n,m}^U + \frac{{\sigma _1^2}}{{\sigma _1^2 + \sigma _2^2}}(\hat h_{CS,n,m}^D - \hat h_{CS,n,m}^U).
\end{equation}
Next, the demodulated communication received symbol is
\begin{equation}\label{equ:d_nm_i}
	\tilde d_{n,m}^i = \frac{{y_{C,n,m}^i}}{{\sqrt {P_t^i} \hat h_{CS,n,m}^{DU}}},
\end{equation} 
where $i = U$ or $D$ are for UL and DL demodulation, respectively; $y_{C,n,m}^i = {({\bf{w}}_{RX}^i)^H}{\bf{y}}_{C,n,m}^i$ is the received data signal after BF, and ${\bf{y}}_{C,n,m}^i$ is given in \eqref{equ:y_C^U}. Based on the ML criterion, the UL and DL communication data can be decoded as 
\begin{equation}\label{equ:d_nm_i_est}
	\hat d_{n,m}^i = \mathop {\arg \min }\limits_{d \in {\Theta _{QAM}}} \left\| {\tilde d_{n,m}^i - d} \right\|_2^2,
\end{equation}
where ${\Theta _{QAM}}$ is the used QAM constellation.

\begin{algorithm}[!t]
	\caption{DUC JCAS Sensing Data Fusion Method}
	\label{DUC_JCAS_fusion_algorithm}
	\KwIn{The sensing results set $\Phi^U$ and $\Phi_1^{DS}$.
		
	}
	\KwOut{The fused DUC JCAS estimation set $\bar \Phi $.}
	\textbf{Step} 1: Count the numbers of points in $\Phi^U$ and $\Phi_1^{DS}$ as $K_1$ and $K_2$, respectively, calculate the normalized distance matrix between $\Phi^U$ and $\Phi_1^{DS}$ by applying \eqref{equ:Z} as ${\bf{Z}}$, and generate a null set $\bar \Phi $.
	
	\textbf{Step} 2: \For{$k$ = 1 to $K_1$} {
		$ind_l = \arg \mathop {\min }\limits_l {\left[ {\bf{Z}} \right]_{k,l}}$\;
		Fuse the $k$th point of $\Phi^U$ with the $ind_l$th point of $\Phi_1^{DS}$ by applying \textbf{Theorem~\ref{Theo:Fusion_criterion}}\;
		Put the fused results into set $\bar \Phi $\;
	}
	\textbf{Step} 3: The remaining points in $\Phi_1^{DS}$ that are not matched, are finally put into $\bar \Phi $.
	
	\Return $\bar \Phi $.
\end{algorithm}

\begin{figure*}[!ht]
	\centering
	\subfigure[Location estimation SMSEs of \textit{cases} 1 and 2.]{\includegraphics[width=0.33\textheight]
		{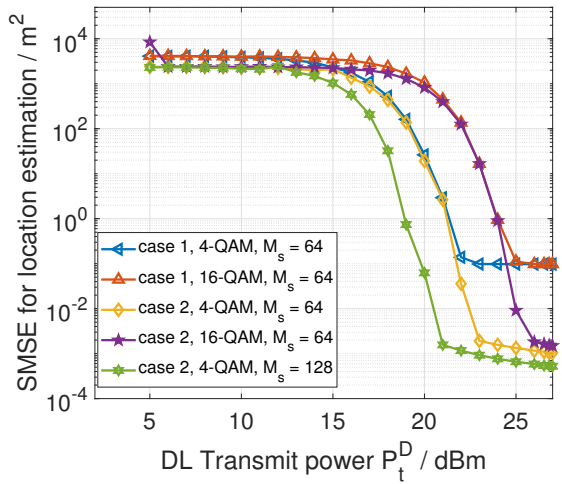}
		\label{figs:SMSE_loc_1234}
	}
	\subfigure[Radial velocity estimation SMSEs of \textit{cases} 1 and 2.]{\includegraphics[width=0.33\textheight]
		{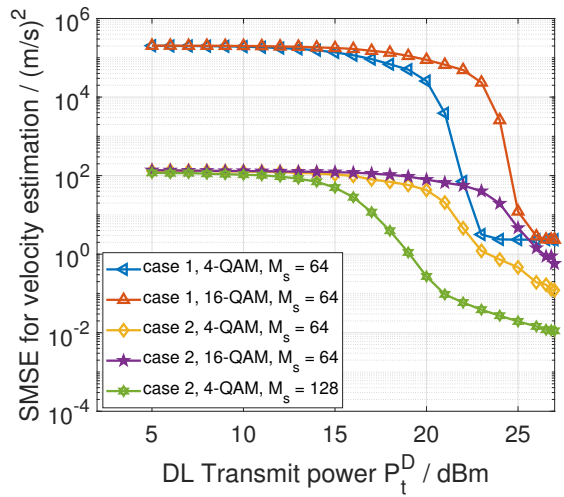}
		\label{figs:SMSE_v_1234}
	}
	\caption{The location and radial velocity estimation SMSEs of \textit{cases} 1 and 2.}
	\label{fig:SMSE_1}
\end{figure*}

\begin{figure*}[!ht]
	\centering
	\subfigure[Location estimation SMSEs of \textit{cases} 3 and 4.]{\includegraphics[width=0.33\textheight]
		{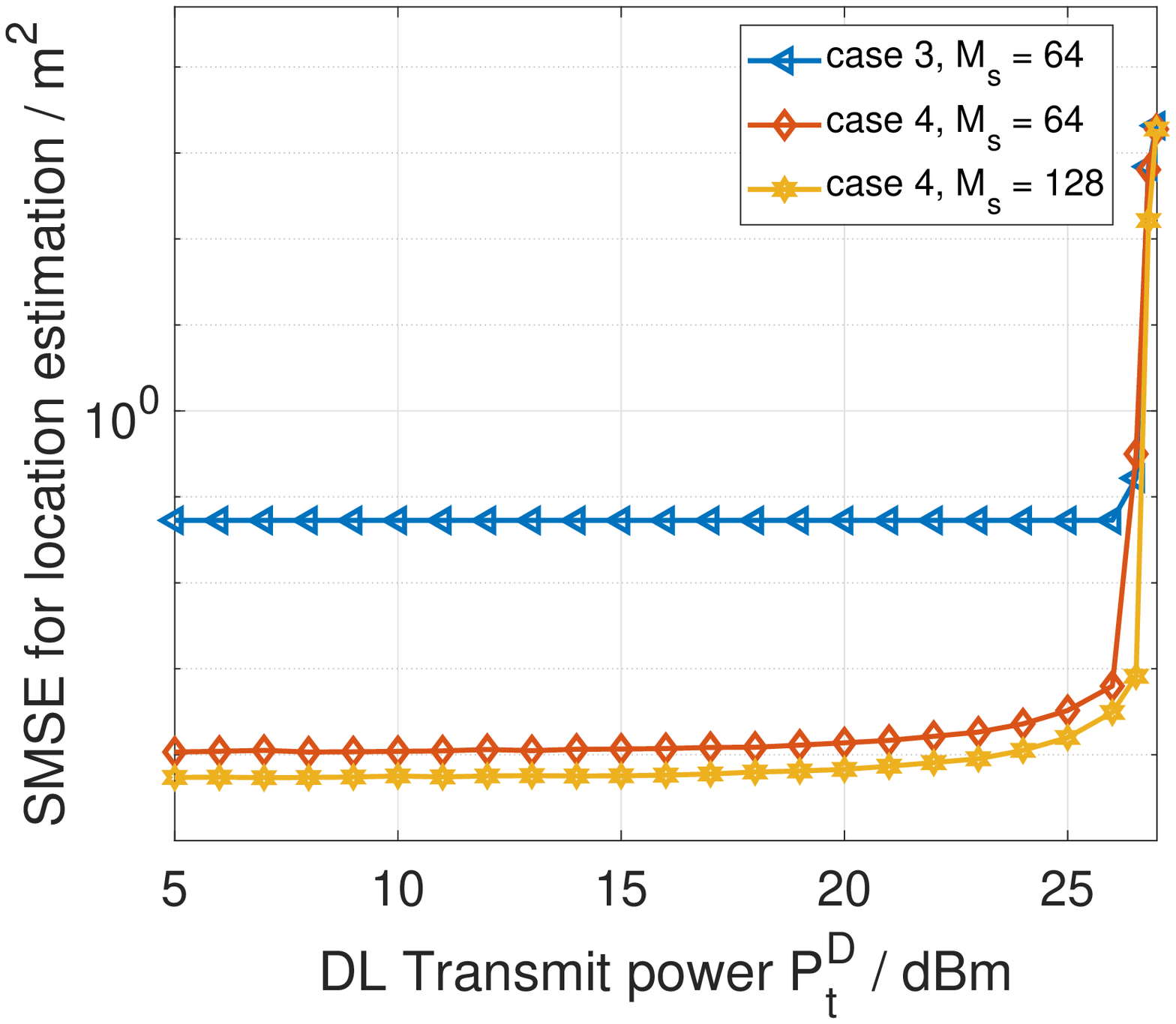}
		\label{figs:SMSE_loc_34} 
	}
	\subfigure[Radial velocity estimation SMSEs of \textit{cases} 3 and 4.]{\includegraphics[width=0.33\textheight]
		{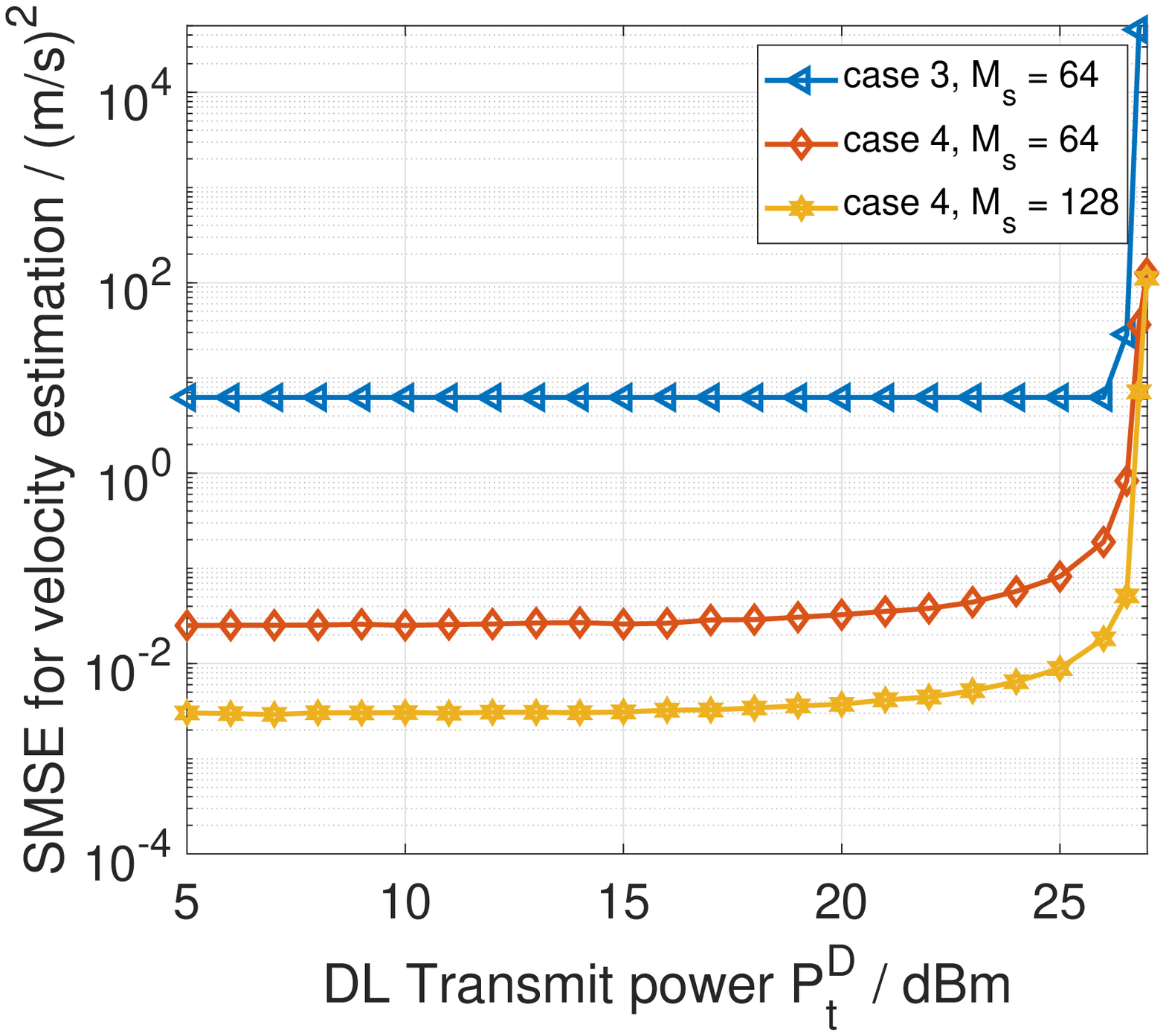}
		\label{figs:SMSE_v_34}
	}
	\caption{The location and radial velocity estimation SMSEs of \textit{cases} 3 and 4.}
	\label{fig:SMSE_2}
\end{figure*}

\section{Simulation Results}\label{sec:Simulation}
\begin{figure*}[!ht]
	\centering
	\subfigure[Location estimation SMSEs.]{\includegraphics[width=0.33\textheight]
		{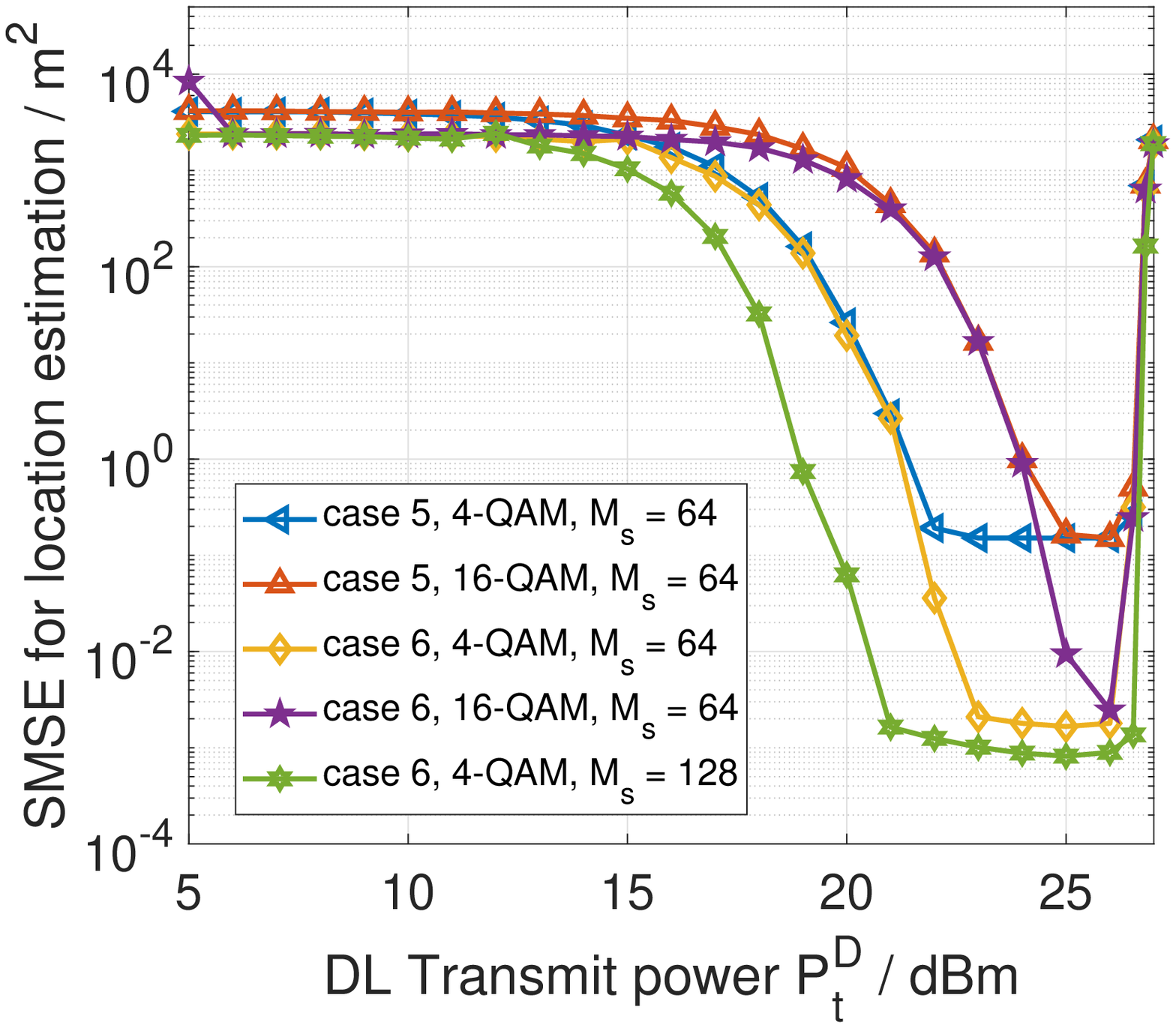}
		\label{fig:SMSE_Ms_location}
	}
	\subfigure[Radial velocity estimation SMSEs.]{\includegraphics[width=0.34\textheight]
		{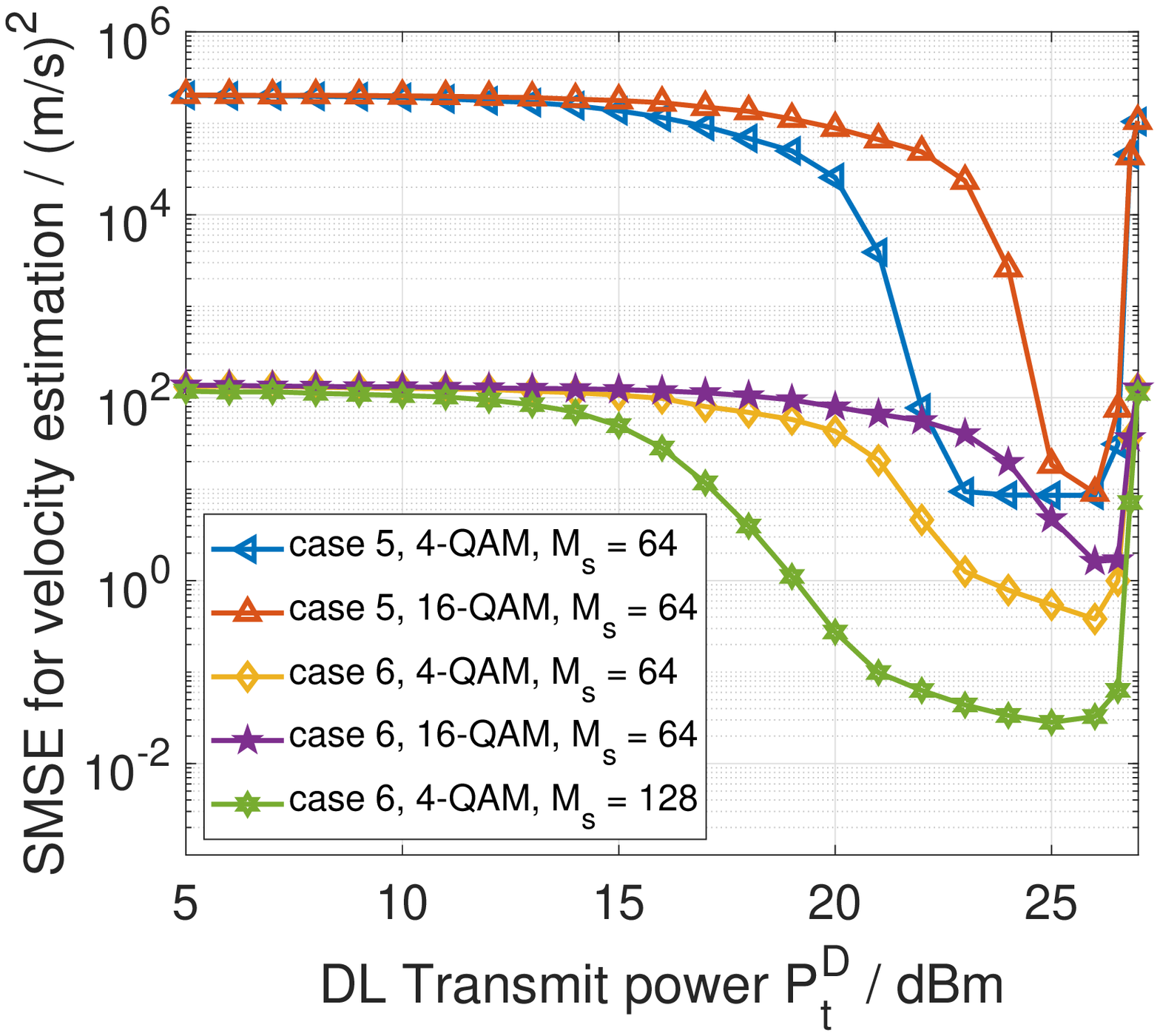}
		\label{fig:SMSE_Ms_vel}
	}
	\caption{The location and radial velocity estimation SMSEs of \textit{cases} 5 and 6 under different QAM  orders and $M_s$.}
	\label{fig:SMSE_Ms}
\end{figure*}
In this section, we present the sensing and communication performance of the proposed DUC JCAS. For comparison, we also plot the sensing and communication performance of the conventional separated UL and DL JCAS, where BS only senses the environment in a single time slot with an on-grid sensing scheme~\cite{Liu2020}. The simulation parameters are listed as follows.

The carrier frequency is set to 63 GHz~\cite{3GPPV2X}, the antenna interval, $d_a$, is half the wavelength, the sizes of antenna arrays of BS and the user are $P_t \times Q_t = 8 \times 8$ and $P_r \times Q_r = 1\times 1$, respectively. The subcarrier interval is $\Delta {f^U} = \Delta {f^D} = \Delta {f} =$ 480 kHz, the subcarrier numbers for UL and DL JCAS are set to $N_c^D = N_c^U = N_c =$ 256, and the bandwidth for JCAS is ${{B  =  }}{N_c}\Delta f = $122.88 MHz. The OFDM symbols used for UL and DL JCAS are set to be the same, i.e., $M_s^U = M_s^D = M_s$.
The variance of the Gaussian noise is $\sigma_N^2 = kFTB = 4.9177\times10^{-12} $ W, where $k = 1.38 \times 10^{-23}$ J/K is the Boltzmann constant, $F = $ 10 is the noise factor, and $T = 290$ K is the standard temperature. The maximum DL and UL transmit power are ${\bar P_t^D}$ = 27 dBm and ${\bar P_t^U}$ = 20 dBm. 
The locations of BS and the user as (50, 4.75, 7) m and (140, 0, 2) m, respectively. The location of the scatterer in DoU is (132, 4.5, 3) m, and the location of the target in DoI is (120, 20, 7) m. Moreover, we set the reflection factors of the targets are $\sigma _{C\beta ,l}^2 = \sigma _{S\beta ,l}^2 = 1$. The velocity of the scatterer in DoI is ($-$40, 0, 0) km/h, and the velocities of the BS and user are (0, 0, 0) m/s. {\color{blue} Note that the above locations and velocities of UE and scatterers are only used for generating some parameters in the simulation setup according to Section~\ref{subsec:JCAS_channel}, and they are unknown to BS. Then, the range, relative velocity, and location of the UE and targets can be estimated by applying the proposed DUC JCAS signal processing scheme.} 
 
The estimation MSEs of range, velocity, and location are defined as the mean values of the squared error of all the estimates, respectively. The velocity is calculated as ${\hat v} = \lambda {\hat f_d}$, where $\hat f_d$ is the estimated Doppler of the target, and $\lambda$ is the wavelength. To measure the sensing performance of estimating all the targets mentioned above, we use the sum of the estimation MSEs of all the targets in specified directions as the sensing performance, which is named the sum of MSEs (SMSE). To simplify the demonstration, we predefine 6 cases, which are listed as follows:
 
\textit{Case} 1: The estimation SMSE of the targets in DoU with conventional separated DL and UL JCAS in~\cite{Liu2020}.

\textit{Case} 2: The estimation SMSE of the targets in DoU with the proposed DUC JCAS.

\textit{Case} 3: The estimation SMSE of the targets in DoI with conventional separated DL and UL JCAS in~\cite{Liu2020}. 

\textit{Case} 4: The estimation SMSE of the targets in DoI with the proposed DUC JCAS. 

\textit{Case} 5: The estimation SMSE of all the targets with conventional separated DL and UL JCAS in~\cite{Liu2020}.

\textit{Case} 6: The estimation SMSE of all the targets with the proposed DUC JCAS.

\subsection{Sensing Performance}
Figs.~\ref{figs:SMSE_loc_1234} and \ref{figs:SMSE_v_1234} present the location and radial velocity estimation SMSEs of \textit{cases} 1 and 2. As $P_t^D$ increases, the sensing SNR in DoU increases, which leads to the decrease in the location and velocity estimation SMSEs of \textit{cases} 1 and 2. Given the same $M_s$ and QAM order, we see that the location and velocity estimation SMSEs of \textit{case} 2 are lower than those of \textit{case} 1 since the proposed DUC JCAS fuses the DL and UL off-grid super-resolution estimation results to enhance the sensing accuracy. Given $M_s = $ 64, the higher QAM order leads to higher SMSEs for both \textit{cases} 1 and 2. This is because the higher QAM order causes larger equivalent noise as shown in \eqref{equ:h_S1_nm} and \eqref{equ:h_S2_nm}. Given the same QAM order, the estimation SMSEs of \textit{case} 2 under the larger $M_s$ become smaller, because more OFDM symbols result in more energy for sensing as shown in~\eqref{equ:autocorrelation} and \eqref{equ:f_r_DS}.

Figs.~\ref{figs:SMSE_loc_34} and \ref{figs:SMSE_v_34} show the location and radial velocity estimation SMSEs of \textit{cases} 3 and 4. The increase of $P_t^D$ leads to the decrease of $P_t^{DS}$ as shown in~\eqref{equ:y_S^DS}, and hence the sensing SNR in DoI decreases, thereby increasing the SMSEs of location and velocity estimations for \textit{cases} 3 and 4. 
Given the same $M_s$, the SMSEs of \textit{case} 4 is lower than \textit{case} 3, benefiting from the off-grid super-resolution estimation ability of DUC JCAS. Moreover, the SMSEs decrease with the growth of $M_s$ for \textit{case} 4 as more energy is accumulated for sensing when more OFDM symbols are used.



Figs.~\ref{fig:SMSE_Ms_location} and \ref{fig:SMSE_Ms_vel} show the location and velocity estimation SMSEs of \textit{cases} 5 and 6 under different QAM orders and $M_s$. The estimation SMSEs of \textit{cases} 5 and 6 are the sum of \textit{cases} 1 and 3, and the sum of \textit{cases} 2 and 4, respectively. Therefore, as $P_t^D$ increases, the location and velocity estimation SMSEs decrease at first, then increase to a large value when $P_t^{DS}$ becomes too small. Given the same QAM order and $M_s$, the SMSEs of \textit{case} 6 are about 20 dB lower than those of \textit{case} 5 because the proposed DUC JCAS can fuse the DL and UL off-grid super-resolution estimation results to enhance sensing performance. Given the same QAM order, the increase of $M_s$ leads to a decrease in estimation SMSEs. This is because the aggregate energy used for sensing increases as $M_s$ increases. The increase of $M_s$ makes the velocity estimation SMSE decrease by more percent than the location estimation SMSE. Specifically, the required $P_t^D$ to achieve the same location estimation SMSE is about 2 dBm lower for \textit{case} 6 with $M_s$ = 128 than that with $M_s$ = 64, while the decrease of required $P_t^D$ is about 3 dBm to achieve the same velocity estimation SMSE. This is because the increase of $M_s$ directly increases the length of symbol time, which leads to the higher Doppler accuracy.

\subsection{Communication Performance}

{\color{blue} 
	In our proposed DUC JACS schemes, BS and UE demodulate the communication data with the refined CSI, $\hat h_{CS,n,m}^{DU}$, as shown in \eqref{equ:SNR_fuse}. By comparison, the conventional separated DL and UL JCAS system uses the CSI estimated in one DLP or ULP frames, i.e., \eqref{equ:y_CS_nm} or \eqref{equ:hCS_D}, to demodulate the communication data. In this subsection, we simulate communications using these two schemes, respectively. The number of trails for generating each result is $N_{cir} = $ 10$^4$.
}

\begin{figure}[!t]
	\centering
	\includegraphics[width=0.34\textheight]{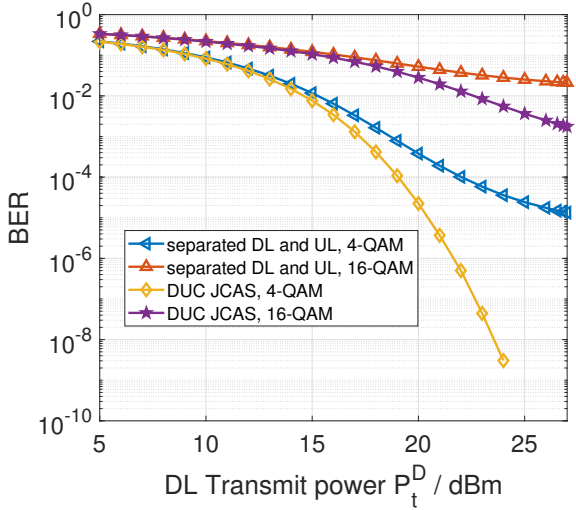}%
	\DeclareGraphicsExtensions.
	\caption{The BERs of the proposed DUC JCAS and the conventional separated DL and UL communication, under 4-QAM and 16-QAM.}
	\label{fig: BER}
\end{figure}

Fig.~\ref{fig: BER} shows BERs of the proposed DUC JCAS and the conventional separated DL and UL communication under 4-QAM and 16-QAM. As $P_t^D$ increases, BER decreases. Given the same QAM order, the BER of the proposed DUC JCAS is lower than that of the conventional communication scheme. This is because DUC JCAS can fuse the estimated CSI in ULP and DLP periods to generate a more accurate CSI. 
Particularly, the BER gap between these two schemes with 16-QAM modulation is larger than that with 4-QAM modulation, because the higher QAM order is more sensitive to CSI estimation errors, leading to the larger BER gap.

\section{Conclusion}\label{sec:conclusion}
In this paper, we propose a DUC JCAS scheme and corresponding DUC JCAS signal processing scheme, which includes a unified UL and DL JCAS sensing scheme and a DUC JCAS fusion method. The unified UL and DL JCAS sensing scheme can achieve off-grid super-resolution estimation for AoA, range, and Doppler with a MUSIC-based sensing processing module. By leveraging the correlation between UL and DL JCAS channels, the DUC JCAS fusion method can distinguish between the sensing results of the user and other dumb scatterers, improving the sensing accuracy. By exploiting the channel reciprocity in the consecutive UL and DL timeslots, the DUC JCAS signal processing scheme can refine the estimated CSI and achieves higher communication reliability.


\begin{appendices}

		\section{Proof of Theorem~\ref{Theo:match}}
		\label{appendix:match}
		
		We note that ${{\bf{A}}_{\bf{r}}}$ and ${{\bf{A}}_{\bf{f}}}$ are composed of exponential functions, i.e., ${e^{ - j2\pi n\Delta f\frac{r_l}{c}}}$ and ${e^{j2\pi m{T_s}f_l}}$. We can obtain that the maximum value of $\| {{{\left[ {{{\bf{a}}_r}\left( r \right)} \right]}^H}{{\bf{A}}_{\bf{r}}}{{\bf{S}}_S}{{\left( {{{\bf{A}}_{\bf{f}}}} \right)}^T}{{\left[ {{{\bf{a}}_f}\left( f \right)} \right]}^*}} \|_2^2$ is $N_c M_s$, which is located at $\left( {r = {r_l}, f = {f_l}} \right)$, ${l = 0,1,\cdots,L - 1}$. Further, since ${\bf{S}}$ is a diagonal matrix with diagonal elements independent of ${{\bf{A}}_{\bf{r}}}$ and ${{{\bf{A}}_{\bf{f}}}}$, when $\left( {r = {r_l}, f = {f_l}} \right)$, $\| {{{[ {{{\bf{a}}_r}\left( r \right)} ]}^H}{{\bf{A}}_{\bf{r}}}{{\bf{S}}_S}{{\left( {{{\bf{A}}_{\bf{f}}}} \right)}^T}{{\left[ {{{\bf{a}}_f}\left( f \right)} \right]}^*}} \|_2^2$ can achieve the local maximum. 
		 
		\section{Proof of Theorem~\ref{Theo:Reference_channel_response}}
		\label{appendix:Reference_channel}
		According to the feature of mono-static active sensing, for ${\bf{H}}_{SU}$, only when ${\bf{p}}_S \in {\Theta }$, BS receives observable echo signals. Otherwise, ${\left( {{{\bf{w}}_{RX}}} \right)^H}{\bf{H}}_{SU}{{\bf{w}}_{TX}} \approx 0$~\cite{richards2010principles}. When ${\bf{p}}_S = {\bf{p}}_{S,\bar k} \in \Theta$, there is also ${\bf{p}}_S \notin { {\{ {{\bf{p}}_{S,k}} \}} |_{k = 0,\cdots,\bar k - 1,\bar k + 1,\cdots,K - 1}}$, we hence have
		\begin{equation}\label{equ:wrHSUTs}
			{( {{{\bf{w}}_{RX}}} )^H}{\bf{H}}_{SU}{{\bf{w}}_{TX}}\!\! \approx\! {b_{S,\bar k}}{( {{{\bf{w}}_{RX}}} )^H}{\bf{a}}( {{\bf{p}}_{S,\bar k}} ){{\bf{a}}^T}( {{\bf{p}}_{S,\bar k}} ){{\bf{w}}_{TX}},
		\end{equation}
		where ${b_{S,\bar k}}$ is a complex value.
		Besides, according to \eqref{equ:H_RS}, we have
		\begin{equation}\label{equ:HRS_expand}
			{({{\bf{w}}_{RX}})^H}{{\bf{H}}_{RS}}{{\bf{w}}_{TX}} = {b_0}{({{\bf{w}}_{RX}})^H}{\bf{a}}({{\bf{p}}_S}){{\bf{a}}^T}({{\bf{p}}_S}){{\bf{w}}_{TX}},
		\end{equation}
		where ${b_0}$ is a complex value.
		By comparing \eqref{equ:wrHSUTs} with \eqref{equ:HRS_expand}, we have 
		\begin{equation}\label{equ:compareHRS_HSU}
			{( {{{\bf{w}}_{RX}}} )^H}{\bf{H}}_{RS}{{\bf{w}}_{TX}} = {k_0}{( {{{\bf{w}}_{RX}}} )^H}{\bf{H}}_{SU}{{\bf{w}}_{TX}},
		\end{equation}
		where ${k_0}$ is a complex value. Specially, when ${\bf{p}}_S \notin \Theta$, ${k_0} \approx 0$. 
		
		Based on \eqref{equ:compareHRS_HSU}, we obtain the conclusions in \textbf{Theorem~\ref{Theo:Reference_channel_response}}.
		
		{\color{blue} 
			\section{Derivation of \eqref{equ:P1} and \eqref{equ:P2}}
			\label{appendix:derivation_P1P2}
			
			From the constraints of the problems \eqref{equ:P1} and \eqref{equ:P2}, we can conclude that ${\bf{w}}_{n,m}^D$ and ${\bf{w}}_{n,m}^{DS}$ are in the nullspaces of ${\bf{H}}_{RS,n,m}^D$ and ${\bf{H}}_{IS,n,m}^D$, respectively. By applying SVD to ${\bf{H}}_{RS,n,m}^D$ and ${\bf{H}}_{IS,n,m}^D$, we can obtain the nullspace bases from the left singular matrices of ${\bf{H}}_{RS,n,m}^D$ and ${\bf{H}}_{IS,n,m}^D$, denoted by ${\bf{U}}_{RS,n,m}^{DN}$ and ${\bf{U}}_{IS,n,m}^{DN}$, respectively. Then, we obtain 
			\begin{equation}\label{equ:w_nm}
				\begin{aligned}
					{\bf{w}}_{n,m}^D &= {\bf{U}}_{RS,n,m}^{DN}{{\bf{m}}_2}, \\
					{\bf{w}}_{n,m}^{DS} &= {\bf{U}}_{IS,n,m}^{DN}{{\bf{m}}_3},
				\end{aligned}
			\end{equation}
			where $\| {{{\bf{m}}_2}} \|_2^2 = \| {{{\bf{m}}_3}} \|_2^2 = 1$. By substituting \eqref{equ:w_nm} into the problems \eqref{equ:P1} and \eqref{equ:P2}, we obtain
			\begin{equation}\label{equ:P3}
				\begin{array}{l}
					\mathop {\max }\limits_{{{\bf{m}}_2}} \| {{{\left( {{{\bf{m}}_2}} \right)}^H}{{\left( {{\bf{U}}_{RS,n,m}^{DN}} \right)}^H}{\bf{H}}_{IS,n,m}^D{\bf{w}}_{TX}^D} \|_2^2\\
					\quad s.t.~\| {{{\bf{m}}_2}} \|_2^2 = 1,
				\end{array}
			\end{equation}
			and
			\begin{equation}\label{equ:P4}
				\begin{array}{l}
					\mathop {\max }\limits_{{{\bf{m}}_3},{{\bf{m}}_1}} \| {{{({{\bf{m}}_3})}^H}{{({\bf{U}}_{IS,n,m}^{DN})}^H}{\bf{H}}_{RS,n,m}^D{\bf{V}}_{C,n,m}^{DN}{{\bf{m}}_1}} \; \|_2^2\\
					\quad\quad s.t.~\| {{{\bf{m}}_1}} \|_2^2 = \left\| {{{\bf{m}}_3}} \right\|_2^2 = 1,
				\end{array}
			\end{equation}
			By applying SVD to ${( {{\bf{U}}_{RS,n,m}^{DN}} )^H}{\bf{H}}_{IS,n,m}^D{\bf{w}}_{TX}^D$ and ${({\bf{U}}_{IS,n,m}^{DN})^H}{\bf{H}}_{RS,n,m}^D{\bf{V}}_{C,n,m}^{DN}$, we obtain 
			\begin{equation}\label{equ:svd_1}
				{( {{\bf{U}}_{RS,n,m}^{DN}} )^H}{\bf{H}}_{IS,n,m}^D{\bf{w}}_{TX}^D = {\bf{U}}_{IS}^D{\bf{\Sigma }}_{IS}^D{( {{\bf{V}}_{IS}^D} )^H},
			\end{equation}
			\begin{equation}\label{equ:svd_2}
				{({\bf{U}}_{IS,n,m}^{DN})^H}{\bf{H}}_{RS,n,m}^D{\bf{V}}_{C,n,m}^{DN} = {\bf{U}}_{RS}^D{\bf{\Sigma }}_{RS}^D{( {{\bf{V}}_{RS}^D} )^H},
			\end{equation}
			where ${\bf{\Sigma }}_{IS}^D$ and ${\bf{\Sigma }}_{RS}^D$ are the real-value diagonal matrices with singular values sorted in the descending order, ${\bf{U}}_{IS}^D$, ${\bf{V}}_{IS}^D$, ${\bf{U}}_{RS}^D$, and ${\bf{V}}_{RS}^D$ are the corresponding right and left singular matrices, respectively, and they are all unitary orthogonal matrices. Therefore, the solutions to \eqref{equ:P3} and \eqref{equ:P4} are
			\begin{equation}\label{equ:M2_M3_M1}
				\begin{aligned}
					{{\bf{m}}_2} = {\left[ {{\bf{U}}_{IS}^D} \right]_{:,1}}, \;
					{{\bf{m}}_3} = {\left[ {{\bf{U}}_{RS}^D} \right]_{:,1}},\;
					{{\bf{m}}_1} = {\left[ {{\bf{V}}_{RS}^D} \right]_{:,1}}.
				\end{aligned}
			\end{equation}
			By applying \eqref{equ:M2_M3_M1} into \eqref{equ:W_TX_DS} and \eqref{equ:w_nm}, we finally obtain \eqref{equ:w_nm_D_DS}.
		}
		
		\section{Proof of Theorem~\ref{Theo:Fusion_criterion}} \label{appendix:FUSION_criterion}
		Since ${\bf{\bar v}} = (1 - \alpha ){{\bf{v}}_1} + \alpha {{\bf{v}}_2}$, we obtain the problem
		\begin{equation}\label{equ:P5}
			\begin{array}{l}
				\mathop {\min }\limits_\alpha  \bar \sigma  = {(1 - \alpha )^2}{\sigma _1^2} + {\alpha ^2}{\sigma _2^2}\\
				s.t. \ 0 < \alpha  < 1.
			\end{array}
		\end{equation}
        As $\frac{{\partial {{\bar \sigma }^2}}}{{{\partial ^2}\alpha }} > 0$, the problem is convex. By solving $\frac{{\partial \bar \sigma }}{{\partial \alpha }} = 0$, we obtain the optimal value of $\alpha$ as $\alpha^*  = \frac{{{\sigma _1^2}}}{{{\sigma _1^2} + {\sigma _2^2}}}$. By substituting $\alpha^* $ into \eqref{equ:P5}, the minimum variance is  $\frac{{{\sigma _1^2}{\sigma _2^2}}}{{{\sigma _1^2} + {\sigma _2^2}}}$. 	
\end{appendices}


%

{\small
	\bibliographystyle{IEEEtran}
	\bibliography{reference}
}
\vspace{-10 mm}
\ifCLASSOPTIONcaptionsoff
  \newpage
\fi

\end{document}